%% file: main.tex
\documentclass[lettersize,journal]{IEEEtran}
\usepackage{amsmath,amsfonts}
\usepackage{algorithmic}
\usepackage{algorithm}
\usepackage{array}
\usepackage[caption=false,font=normalsize,labelfont=sf,textfont=sf]{subfig}
\usepackage{textcomp}
\usepackage{stfloats}
\usepackage{url}
\usepackage{verbatim}
\usepackage{color,soul}
\usepackage{graphicx}
\usepackage{hyperref}
\usepackage{cite}
\usepackage[nolist]{acronym}
\usepackage{mathtools}
\usepackage{amssymb}
\usepackage{stfloats}% <-- added
\usepackage{tikzit} % Step 1
\usepackage{titlesec}

\input{./figures/sample.tikzstyles} % Step 2

\acrodefplural{HPA}[HPAs]{high power amplifiers}

\input{acronyms.tex}

\DeclareMathOperator{\trace}{tr}

\DeclareMathOperator{\ve}{vec}
\DeclareMathOperator{\Real}{Re}
\DeclareMathOperator{\Imag}{Im}
\DeclareMathOperator{\dB}{dB}

\DeclareMathOperator{\NMSE}{NMSE}
\DeclareMathOperator{\GHz}{GHz}

\DeclareMathOperator*{\argmin}{arg\,min}

\newcommand\myeqa{\mathrel{\overset{\makebox[0pt]{\mbox{\normalfont\tiny\sffamily (a)}}}{=}}}
\newcommand\myeqb{\mathrel{\overset{\makebox[0pt]{\mbox{\normalfont\tiny\sffamily (b)}}}{=}}}
\newcommand\myeqc{\mathrel{\overset{\makebox[0pt]{\mbox{\normalfont\tiny\sffamily (c)}}}{=}}}
\newcommand\myeqd{\mathrel{\overset{\makebox[0pt]{\mbox{\normalfont\tiny\sffamily (d)}}}{=}}}
\newcommand\myeqe{\mathrel{\overset{\makebox[0pt]{\mbox{\normalfont\tiny\sffamily (e)}}}{=}}}
\newcommand\myeqf{\mathrel{\overset{\makebox[0pt]{\mbox{\normalfont\tiny\sffamily (f)}}}{=}}}

\newcommand\mygeqa{\mathrel{\overset{\makebox[0pt]{\mbox{\normalfont\tiny\sffamily (a)}}}{\geq}}}

\newcommand\mygeqd{\mathrel{\overset{\makebox[0pt]{\mbox{\normalfont\tiny\sffamily (d)}}}{\geq}}}
\newcommand\mygeqe{\mathrel{\overset{\makebox[0pt]{\mbox{\normalfont\tiny\sffamily (e)}}}{\geq}}}
\newcommand\mygeqf{\mathrel{\overset{\makebox[0pt]{\mbox{\normalfont\tiny\sffamily (f)}}}{\geq}}}

\newcommand\myleqa{\mathrel{\overset{\makebox[0pt]{\mbox{\normalfont\tiny\sffamily (a)}}}{\leq}}}

\newcommand\myleqd{\mathrel{\overset{\makebox[0pt]{\mbox{\normalfont\tiny\sffamily (d)}}}{\leq}}}

\newcommand\myapproxeqb{\mathrel{\overset{\makebox[0pt]{\mbox{\normalfont\tiny\sffamily (b)}}}{\simeq}}}

\newcommand\myapproxeqe{\mathrel{\overset{\makebox[0pt]{\mbox{\normalfont\tiny\sffamily (e)}}}{\simeq}}}

\usepackage{xcolor,cite,etoolbox}
\makeatletter 
\pretocmd\@bibitem{\color{black}\csname keycolor#1\endcsname}{}{\fail}
\newcommand\citecolor[1]{\@namedef{keycolor#1}{\color{blue}}}
\makeatother

\usepackage{etoolbox} % For patching commands

\makeatletter
\newcommand{\mypcite}[1]{%
  \begingroup
  \leavevmode
  \def\@cite##1##2{[##1]}% Redefine \@cite to remove space
  (\cite{#1})%
  \endgroup
}
\makeatother

\begin{document}

\title{Mutual Information Based Pilot Design for ISAC}

\author{Ahmad Bazzi,  Marwa Chafii 
\thanks{
Ahmad Bazzi and Marwa Chafii is with Engineering Division, New York University (NYU) Abu Dhabi, 129188, UAE and NYU WIRELESS,
NYU Tandon School of Engineering, Brooklyn, 11201, NY, USA (email: \href{ahmad.bazzi@nyu.edu}{ahmad.bazzi@nyu.edu}, \href{marwa.chafii@nyu.edu}{marwa.chafii@nyu.edu}).}
}

\markboth{to appear in IEEE Transactions on Communications, 2025}%
{Shell \MakeLowercase{\textit{et al.}}: A Sample Article Using IEEEtran.cls for IEEE Journals}

\IEEEpubid{}

\maketitle

\setlength{\abovedisplayskip}{3pt}
\setlength{\belowdisplayskip}{2pt}
 \setlength{\abovedisplayskip}{1pt}
 \setlength{\belowdisplayskip}{2pt}
 \titlespacing{\section}{0pt}{*0.05}{*0.1}
 \titlespacing{\subsection}{0pt}{*0.05}{*0.1}
 \setlength{\textfloatsep}{2pt} 
 \setlength{\floatsep}{1pt}      
 \setlength{\intextsep}{1pt}

\begin{abstract}
The following paper presents a novel orthogonal pilot design dedicated for \textcolor{black}{integrated sensing and communications (ISAC)} systems performing multi-user communications and target detection. After careful characterization of both sensing and communication metrics based on mutual information (MI), we propose a multi-objective optimization problem (MOOP) tailored for pilot design, dedicated for simultaneously maximizing both sensing and communication MIs. Moreover, the MOOP is further simplified to a single-objective optimization problem, which characterizes trade-offs between sensing and communication performances. Due to the non-convex nature of the optimization problem, we propose to solve it via the projected gradient descent method on the Stiefel manifold. Closed-form gradient expressions are derived, which enable execution of the projected gradient descent algorithm. Furthermore, we prove convergence to a fixed orthogonal pilot matrix. Finally, we demonstrate the capabilities and superiority of the proposed pilot design, and corroborate relevant trade-offs between sensing MI and communication MI. 
In particular, significant signal-to-noise ratio (SNR) gains for communication are reported, while re-using the same pilots for target detection with significant gains in terms of probability of detection for fixed false-alarm probability. Other interesting findings are reported through simulations, such as an \textit{information overlap} phenomenon, whereby the fruitful ISAC integration can be fully exploited. 
\end{abstract}


\section{Introduction}
\label{sec:introduction}
\input{sections/introduction.tex}

\section{System Model}
\label{sec:system-model}
\input{sections/system-model.tex}

\section{Performance Metrics}
\label{sec:performance-metrics}
\input{sections/performance-metrics.tex}

\section{Optimization Framework for ISAC Pilot Design}
\label{sec:opt-probs}
\input{sections/optimization-problems.tex}

\section{ISAC Pilot Design by Projected Gradient Descent}
\label{sec:opt-01}
\input{sections/optimization01.tex}

\section{\textcolor{black}{On the adopted \ac{MI} metrics: Connections \& Motivation}}
\label{sec:motivation}
\textcolor{black}{\input{sections/motivation.tex}}

\section{Simulation Results}
\label{sec:simulations}
\input{sections/simulation-results.tex}

\section{Conclusions and Research Directions}
\label{sec:conclusions}
\input{sections/conclusions.tex}

\appendices
\label{sec:appendix}
\section{Expression of $\mathcal{M}_k^{\tt{comm}}(\pmb{\Phi})$}
\label{app:MI-comm}
\input{sections/proof-of-Mcomm}

\section{Expression of $\mathcal{M}^{\tt{sense}}(\pmb{\Phi})$}
\label{app:MI-sense}
\input{sections/proof-of-Msense}
\section{Expression of $\nabla_{\pmb{\Phi}} \mathcal{M}_k^{\tt{comm}}(\pmb{\Phi})$}
\label{app:dMI-comm}
\input{sections/proof-of-dMcomm}

\section{Expression of $\nabla_{\pmb{\Phi}_l} \mathcal{M}^{\tt{sense}}(\pmb{\Phi})$}
\label{app:dMI-sense}
\input{sections/proof-of-dMsense}

\section{Proof of Theorem 1}
\label{app:convergence-analysis}
\input{sections/convergence-analysis-proof}

\section{\textcolor{black}{Proof of Theorem 2}}
\label{app:lower-bound-Mcomm}
\textcolor{black}{\input{Actions/lower-bound-on-Mcomm}}

\section{\textcolor{black}{Proof of Theorem 3}}
\label{app:Stein-on-Msense}
\textcolor{black}{\input{Actions/stein-on-m-sense}}

\bibliographystyle{IEEEtran}
\bibliography{refs2}

\vfill

\end{document}

%% file: figures/sample.tikzstyles
\tikzstyle{circle}=[fill={rgb,255: red,229; green,123; blue,255}, draw=black, shape=circle, minimum width=0.01cm, minimum height=0.01cm]
\tikzstyle{optimal}=[fill={rgb,255: red,87; green,188; blue,255}, draw=black, shape=circle]
\tikzstyle{circle2}=[fill={rgb,255: red,255; green,203; blue,82}, draw=black, shape=circle]

\tikzstyle{Stiefel}=[fill={rgb,255: red,189; green,130; blue,255}, draw=black, -]
\tikzstyle{new edge style 0}=[->, fill=none, draw={rgb,255: red,0; green,128; blue,128}]
\tikzstyle{new edge style 1}=[-, fill=none, draw={rgb,255: red,232; green,232; blue,232}]
\tikzstyle{dashed elipse}=[-, dashed, fill={rgb,255: red,193; green,205; blue,255}, opacity=0.5]
\tikzstyle{dashed curve}=[-, dashed]
\tikzstyle{dashed curve 2}=[-, dashdotted]
\tikzstyle{dotted curve}=[-, dotted]
\tikzstyle{arrow black}=[->]

%% file: acronyms.tex
\begin{acronym}
	\acro{SI}{self interference}
    \acro{MIMO}{multiple-input, multiple-output}
    \acro{AWGN}{additive white Gaussian noise}
    \acro{D2D}{device-to-device}
    \acro{RCS}{radar cross-section}
    \acro{ISAC}{integrated sensing and communication}
    \acro{DFRC}{dual-functional radar and communication}
    \acro{AoA}{angle of arrival}
    \acro{ToA}{time of arrival}
    \acro{EVM}{error vector magnitude}
    \acro{CPI}{coherent processing interval}
    \acro{eMBB}{enhanced mobile broadband}
    \acro{URLLC}{ultra-reliable low latency communications}
    \acro{mMTC}{massive machine type communications}
	\acro{QCQP}{quadratic constrained quadratic programming}
	\acro{BnB}{branch and bound}
	\acro{SER}{symbol error rate}
	\acro{LFM}{linear frequency modulation}
	\acro{BS}{base station}
	\acro{UE}{user equipment}
	\acro{CS}{compressed sensing}
	\acro{PR}{passive radar}
	\acro{LMS}{least mean squares}
	\acro{NLMS}{normalized least mean squares}
	\acro{RIS}{reconfigurable intelligent surface}
	\acro{RISs}{reconfigurable intelligent surfaces}
	\acro{IRS}{intelligent reflecting surface}
	\acro{IRSs}{intelligent reflecting surfaces}
	\acro{LISA}{large intelligent surface/antennas}
	\acro{LISs}{large intelligent surfaces}
	\acro{OFDM}{orthogonal frequency-division multiplexing}
	\acro{EM}{electromagnetic}
	\acro{ISMR}{integrated sidelobe to mainlobe ratio}
	\acro{MSE}{mean squared error}
	\acro{SNR}{signal-to-noise ratio}
	\acro{SRP}{successful recovery probability}
	\acro{CDF}{cumulative distribution function}
	\acro{ULA}{uniform linear array}
	\acro{RSS}{received signal strength}
	\acro{SU}{single user}
	\acro{MU}{multi-user}
	\acro{DL}{downlink}
	\acro{i.i.d}{independently and identically distributed}
	\acro{UL}{uplink}
	\acro{LoS}{line of sight}
	\acro{DFT}{discrete Fourier transform}
    \acro{HPA}{high power amplifier}
    \acro{IBO}{input power back-off}
    \acro{MIMO}{multiple-input, multiple-output}
    \acro{PAPR}{peak-to-average power ratio}
    \acro{FD}{full duplex}
    \acro{D2D}{device-to-device}
    \acro{ISAC}{integrated sensing and communication}
    \acro{DFRC}{dual-functional radar and communication}
    \acro{AoA}{angle of arrival}
    \acro{AoD}{angle of departure}
    \acro{ToA}{time of arrival}
    \acro{EVM}{error vector magnitude}
    \acro{CPI}{coherent processing interval}
    \acro{eMBB}{enhanced mobile broadband}
    \acro{URLLC}{ultra-reliable low latency communications}
    \acro{mMTC}{massive machine type communications}
	\acro{QCQP}{quadratic constrained quadratic programming}
	\acro{BnB}{branch and bound}
	\acro{SER}{symbol error rate}
	\acro{LFM}{linear frequency modulation}
	\acro{ADMM}{alternating direction method of multipliers}
	\acro{CCDF}{complementary cumulative distribution function}
	\acro{MISO}{multiple-input, single-output}
	\acro{CSI}{channel state information}
	\acro{LDPC}{low-density parity-check}
	\acro{BCC}{binary convolutional coding}
	\acro{IEEE}{Institute of Electrical and Electronics Engineers}
	\acro{ULA}{uniform linear antenna}
	\acro{B5G}{beyond 5G}
	\acro{MOOP}{multi-objective optimization problem}
	\acro{ML}{maximum likelihood}
	\acro{FML}{fused maximum likelihood}
	\acro{RF}{radio frequency}
	\acro{AM/AM}{amplitude modulation/amplitude modulation}
	\acro{AM/PM}{amplitude modulation/phase modulation}
	\acro{BS}{base station}
	\acro{SCA}{successive convex approximation}
	\acro{MM}{majorization-minimization}
	\acro{LNCA}{$\ell$-norm cyclic algorithm}
	\acro{OCDM}{orthogonal chirp-division multiplexing}
	\acro{TR}{tone reservation}
	\acro{COCS}{consecutive ordered cyclic shifts}
	\acro{LS}{least squares}
	\acro{FIM}{Fisher information matrix}
	\acro{CVE}{coefficient of variation of envelopes}
	\acro{BSUM}{block successive upper-bound minimization}
	\acro{ICE}{iterative convex enhancement}
	\acro{SOA}{sequential optimization algorithm}
	\acro{BCD}{block coordinate descent}
	\acro{SINR}{signal to interference plus noise ratio}
	\acro{MICF}{modified iterative clipping and filtering}
	\acro{PSL}{peak side-lobe level}
	\acro{SDR}{semi-definite relaxation}
	\acro{KL}{Kullback-Leibler}
	\acro{SCNR}{signal-to-clutter-plus-noise ratio}
	\acro{ADSRP}{alternating direction sequential relaxation programming}
	\acro{MUSIC}{MUltiple SIgnal Classification}
	\acro{EVD}{eigenvalue decomposition}
	\acro{SVD}{singular value decomposition}
	\acro{3GPP}{3rd Generation Partnership Project}
	\acro{NLoS}{non line of sight}
	\acro{PSK}{phase shift keying}
	\acro{i.i.d}{independent and identically distributed}
	\acro{PDF}{probability density function}
	\acro{GMM}{Gaussian mixture model}
	\acro{MI}{mutual information}
	\acro{MOOP}{multi-objective optimization problem}
	\acro{RSC}{restricted strongly convex}
	\acro{RSS}{restricted strongly smooth}
	\acro{ROC}{receiver operating characteristic}
	\acro{NMSE}{normalized mean square error}
	\acro{MMSE}{minimum mean square error}
	\acro{SER}{symbol error rate}
	\acro{QAM}{quadrature amplitude modulation}
	\acro{ZF}{zero-forcing}
	\acro{B5G}{beyond 5G}
	\acro{IoT}{internet of things}
	\acro{UAV}{unmanned aerial vehicle}
	\acro{DAM}{delay alignment modulation}
	\acro{ISR}{integrated sidelobe ratio}
	\acro{CRB}{Cram\'er-Rao bound}
	\acro{RCC}{radar-communication coexistence}
	\acro{NSP}{null space projection}
	\acro{OTFS}{orthogonal time frequency space}
	\acro{STARS}{simultaneously transmitting and reflecting surface}
	\acro{AM}{arithmetic mean}
	\acro{GM}{geometric mean}
\end{acronym}

%% file: sections/introduction.tex
\IEEEPARstart{T}{\lowercase{o}} address the spectrum constraint resulting from the growing demand of wireless devices, \ac{ISAC} systems \cite{10041914,9705498,9585321,9540344,10049816} have recently emerged as a key enabler to solve the ever-growing spectrum congestion problem, thus attracting both the interest of academia and industry. 
\ac{ISAC} is currently presented as one strategy to reducing this problem by co-designing wireless communication functions and radar sensing on a single hardware platform and sharing the spectrum of both radar and communication systems, thus improving band-utilization efficiency.
Therefore, radar sensing and communication tasks are carried out through a unified platform and a common radio waveform at the same time, over the same frequency band, utilizing the same antennas.
It is worth noting that both sensing and communication have been treated as two separate fields since the $1900$s. Then, radar waveforms performing communications have appeared, and vice-versa. For example, in $1967$, pulse interval modulation was introduced to modulate communication data onto a pulse-based radar waveform, and \ac{OFDM}-based sensing has been utilized to perform sensing with \ac{OFDM} \cite{sit20132d}. 

Earlier research efforts were dedicated towards spectrum co-existence, also referred to as \ac{RCC} \cite{8828016,8828023,8094973}, whereby both radar and communication co-exist at the price of deliberate interference caused by one sub-system onto the other. Therefore, interference cancellation methods have been designed to address the \ac{RCC} interference issue, such as \textcolor{black}{null space projection}\cite{6503914}. 
Unfortunately, allocating different frequency bands for these systems is neither practical nor sustainable, since both sensing and communication  systems are requesting additional resources.
Although being long thought of as two separate domains, sensing and communication are in fact, intimately entangled from an information-theoretic perspective \cite{10147248}. 
To realize the goal of \ac{ISAC} systems, however, various obstacles must be overcome.
For example, this integration complicates the design of signal waveforms, resource and hardware allocation, and network operation. All of this motivates the search for innovative approaches to these issues in order to allow the advantages of the \ac{ISAC} system in real-world deployments for sensing accuracy and high-rate communications.

\textcolor{black}{\input{Actions/R3C1_rewrite}}

Nevertheless, most of the aforementioned designs focus only on waveform and beamforming designs targeting capacity, spectral efficiency, or sum-rate maximization without addressing channel estimation concerns in the context of \ac{ISAC}. Therefore, we find it crucial to provide pilots, that are good for both channel estimation and target detection. To this end, this paper considers \textcolor{black}{\ac{ISAC}} \ac{BS} pilot design with flexible \ac{ISAC} trade-offs, intended for \ac{DL} communication users, while listening to the backscattered echo of the transmitted pilot. Meanwhile, the users utilize the pilot for channel estimation, for equalization and decoding purposes. 
The design aims at maximizing mutual information, for both communication and sensing, on the Stiefel manifold in order to preserve the orthogonality constraint on the pilots. A design parameter is highlighted that trades off communication and sensing performances. Towards this \ac{ISAC} design, we are faced with a non-convex optimization problem, which can be efficiently and reliably solved via an appropriate projected gradient descent algorithm.  It is worth emphasizing that our results show that these pilots can be effectively utilized for channel estimation at the user’s end, while the \textcolor{black}{\ac{ISAC}} \ac{BS} can exploit the echo of the same pilot to detect targets.
To that purpose, we have summarized our main contributions below.
\input{Actions/contributions.tex}
Furthermore, we unveil some important insights, i.e.
\begin{itemize}
	\item The number of transmit antennas and the number of pilot symbols have profound impact on the \ac{ISAC} \ac{MI} achievable tradeoffs, allowing a flexible design of pilots, even in the presence of clutter. 
	\item The \ac{ISAC} \ac{MI} performance is influenced by the target location, where the closer the target approaches the mean of communication user \ac{AoA}, the pointier the Pareto frontier becomes. In particular, as the target approaches the communication user, the \ac{ISAC} \ac{MI} frontier is pushed outwards towards the utopia point. We term this phenomenon as the \textit{information overlap} phenomenon, whereby the sensing and communication channels share common information. Consequently, using the same orthogonal pilot matrix, which is designed based on the proposed method, allows to simultaneously achieve better sensing and communication MI performance. 
	\item The \ac{ROC} performance of an optimized orthogonal pilot matrix for communication is better than an non-optimized orthogonal pilot matrix. The \ac{ROC} performance can be further improved through a design parameter that controls the priority of sensing over communication.
	\item For multi-user communications, the \ac{SER} performance utilizing the channel estimates aided by the generated pilot matrices exhibit gains as high as $1.6\dB$ for an \ac{SER} level of $10^{-4}$. This gain can be controlled through an \ac{ISAC} design parameter trading off sensing for communications. Furthermore, \ac{SNR} gains of about $6\dB$ are reported for \ac{NMSE} performance of channel estimation.    
	\item \textcolor{black}{\input{Actions/insight-convergence}}
\end{itemize}

The detailed structure of the following paper is given as follows. 
In Section \ref{sec:system-model}, we introduce the \ac{ISAC} system model
Section \ref{sec:performance-metrics} introduces mutual information metrics for the problem of pilot design, which enable us to optimize the pilots for both sensing and communication tasks.
 In Section \ref{sec:opt-probs}, an optimization framework for \ac{ISAC} pilot design is designed. 
Moreover, Section \ref{sec:opt-01} describes a method to solve the pilot design problem in an iterative fashion, and its convergence property is described. 
 Section \ref{sec:simulations} provides numerical results to verify our analysis before concluding the paper in Section \ref{sec:conclusions}.
\textbf{Notation}:
Upper-case and lower-case boldface letters denote matrices and vectors, respectively. 
$(.)^T$, $(.)^*$ and $(.)^H$ represent the transpose, the conjugate and the transpose-conjugate operators. 
The statistical expectation is denoted as \textcolor{black}{$\mathbb{E}[\cdot]$}. 
The $\ell_2$ norm of a vector $\pmb{x}$ is denoted as $\Vert \pmb{x} \Vert$. 
The matrix $\pmb{I}_N$ is the identity matrix of size $N \times N$. 
The zero-vector is $\pmb{0}$. 
For matrix indexing, the $(i,j)^{th}$ entry of matrix $\pmb{A}$ is denoted by $[\pmb{A}]_{i,j}$ and its $j^{th}$ column is denoted as $[\pmb{A}]_{:,j}$. 
\textcolor{black}{\input{Actions/define_vec}}
\textcolor{black}{\input{Actions/C3o3_definenormaldensity}}

%% file: Actions/R3C1_rewrite.tex
%From a waveform design perspective, many efforts have been invested in optimizing waveforms that are deemed suitable for both sensing and communications. 
The integration of sensing and communication into a unified system has opened up new possibilities for improving efficiency and enabling advanced applications. One critical avenue of research has been waveform design, where the goal is to develop waveforms capable of balancing the requirements of sensing and communication. For instance, waveforms have been tailored to minimize multi-user communication interference while preserving essential chirp characteristics with configurable \ac{PAPR} \cite{10061453}. In other cases, full-duplex waveform frameworks utilize the idle time of pulsed radar systems to enable seamless communication transmissions, ensuring efficient resource utilization \cite{9724187}. Sparse vector coding techniques have also been explored to design waveforms with low sidelobes and guaranteed communication performance, addressing the challenges of coexisting functionalities \cite{9695365}. As far as waveform sidelobe is concerned, it was recently proven \cite{liu2024ofdm} that the \ac{OFDM} waveform achieves the lowest sidelobe level at \textit{each lag} of the so-called periodic auto-correlation function for sub-Gaussian constellations, e.g. \ac{QAM} and \ac{PSK} type constellations. On the other hand, it was also proven \cite{liu2024ofdm} that for super-Gaussian constellations single-carrier waveforms with cyclic prefix is the waveform that achieves the lowest sidelobes.
Beyond waveform design, beamforming plays an important role in further enabling \ac{ISAC} systems to achieve joint sensing and communication objectives. For example, an approach is to optimize beamforming matrices to reduce the outage \ac{SINR} probability while maximizing radar output power in the Bartlett sense \cite{10018908}, which can promote \ac{ISAC} applications when imperfect \textcolor{black}{channel state information} is known, due to some estimation errors. 
A weighted combination of the sum rate and the \ac{CRB} was optimized for transmit beamforming design in \cite{Wu2024}, under an available transmit power budget.
From interference management perspective, joint transmit and receive beamformers have proven effective in \textcolor{black}{beyond $5$G} cellular \textcolor{black}{internet of things} scenarios, highlighting the interplay between interference mitigation and \ac{ISAC} performance \cite{9206051}. Further advancements integrate beamforming with dynamic elements such as \ac{UAV} trajectory planning to maximize system performance while adhering to beampattern requirements \cite{9916163}. In other instances, beamformers have been designed to optimize communication \ac{SNR} while maintaining \textcolor{black}{integrated sidelobe ratio} guarantees for radar sensing \cite{10105893}. Additionally, deep learning techniques have shown promise for tracking sensing parameters in vehicular networks, adding a layer of intelligence to \ac{ISAC} operations \cite{9492131}. Theoretical analysis have provided insights into the trade-offs and limitations of \ac{ISAC} systems. Frameworks that assess detection and false-alarm probabilities offer valuable tools for performance evaluation \cite{10124135}, while other studies quantify trade-offs between \ac{CRB} and communication capacity, offering a deeper understanding of \ac{ISAC} system capabilities \cite{10147248}. The evolution of \ac{ISAC} is also being shaped by its integration with emerging technologies, including holographic \ac{MIMO} \cite{9724245}, massive \ac{MIMO} \cite{9898900}, \textcolor{black}{reconfigurable intelligent surface} \cite{bazzi2022ris}, \textcolor{black}{simultaneously transmitting and reflecting surfaces} \cite{10050406}, and \ac{OFDM} \cite{9967989}. These combinations extend the reach and applicability of \ac{ISAC}, positioning it as a cornerstone for future wireless systems.

%% file: Actions/contributions.tex
\begin{itemize}
%	  \item \textbf{Gaussian mixture modeling for \ac{ISAC}}. Our work models the communication channel as a \ac{GMM}, which is intended to approximate the non-Gaussian noise in wireless communication channels. Such a model is well-justified in transmissions where communication users are spatially dispensed following a non-uniform distribution. This paper is the first to push forward \ac{ISAC} boundaries and promote \ac{GMM} communication modeling. 
	\item \input{Actions/contribution-1}
	\item \input{Actions/contribution-2}
	\item \input{Actions/contribution-3}
\end{itemize}

%% file: Actions/contribution-1.tex
\textcolor{black}{\textbf{Unified \ac{MI} metric for pilot design framework}}. 
	\textcolor{black}{We derive a unified \ac{MI} metric that enable pilots to perform target detection, while reliably utilized for channel estimation tasks for \textcolor{black}{multi-user} communications.}
		\textcolor{black}{In particular, a sensing \ac{MI} metric is proposed to optimize target detection performance, while a communication \ac{MI} metric is proposed to aid with channel estimation tasks. In this regard, a \ac{MOOP} is devised to unify the proposed \ac{MI} metrics into an \ac{ISAC} \ac{MI} metric.}
	\textcolor{black}{\input{Actions/C3o1_clarify.tex}}
%	Furthermore, an \ac{MOOP} is designed with the intention of pilot optimization, which are capable of performing simulateneous channel estimation and target detection via the same pilot waveform. 
%Indeed, the \textcolor{black}{\ac{ISAC}} \ac{BS} can leverage the backscattered optimized pilot waveform to perform target detection via the optimal detector. Meanwhile, communication users can use the same pilot waveform to perform reliable channel estimation, followed by channel equalization and, eventually, data decoding. The framework offers \ac{ISAC} trade-offs, thus enabling orthogonal pilot generation with the ability to trade communication performance with sensing, and vice-versa. 
%\textcolor{black}{Thanks to this framework, orthogonal pilot generation can be also utilized  trade-offs between communication and sensing performance within the ISAC system.}

%% file: Actions/C3o1_clarify.tex
Usually, \ac{ISAC} waveforms are designed with the purpose of guaranteeing reliable communication under some constraints on sensing metrics. 
Unlike this approach, the proposed framework designs orthogonal pilot waveforms	 with the purpose of aiding the  communication channel estimation, which we show in Section \ref{sec:motivation}, has positive reverberation effects on the communication capacity.
Moreover, the same pilots are also optimized for target detection.

%% file: Actions/contribution-2.tex
\textbf{Non-convex optimization via projected gradient descent}. Being confronted with a \ac{MOOP}, we first scalarize the problem, which gives rise to single-objective, but non-convex optimization problem, which is to be solved over the Stiefel manifold. We derive a projected gradient descent algorithm tailored for \ac{ISAC} orthogonal pilot waveforms.

%% file: Actions/contribution-3.tex
\textcolor{black}{\input{Actions/C301_additional_contr.tex}} 
%	\item \textbf{Extensive simulation results}. In order to highlight the various benefits of the proposed orthogonal pilot design and the potential of the designed projected gradient descent method in both multi-user communications and radar sensing, we present extensive simulation results demonstrating the potential, superiority and fruitful \ac{ISAC} tradeoffs of the proposed scheme.

%% file: Actions/C301_additional_contr.tex
\textbf{Connections with existing sensing and communication metrics.} We show how the proposed \ac{MI}-based metrics for both communications and sensing are connected with classical performance metrics.
In particular, we show that the communication \ac{MI}-based metric, although intended for channel estimation purposes, positively influences the worst-case channel capacity, which improves communication performance.
Moreover, the proposed sensing \ac{MI}-based metric is asymptotically connected to the detection probability of the most powerful detection test, according to Neyman-Pearson criterion, under a fixed false alarm probability. 

%% file: Actions/insight-convergence.tex
 For sake of completeness, we prove that the proposed projected gradient method approach can always converge to a stable pilot solution.
 In this regard, the convergence properties of the algorithm are derived.

%% file: Actions/define_vec.tex
Also, $\operatorname{vec}$ is the vectorization operator, which vectorizes a matrix by stacking its columns, one on top of the other.

%% file: Actions/C3o3_definenormaldensity.tex
The density $f_{\mathcal{CN}}(\pmb{x} \vert \pmb{\mu},\pmb{\Sigma})$ represents the probability density function of the complex Gaussian random variable $\pmb{x}$ following mean $\pmb{\mu}$ and covariance $\pmb{\Sigma}$.

%% file: sections/system-model.tex
\begin{figure}[!t]
\centering
\includegraphics[width=3.25in]{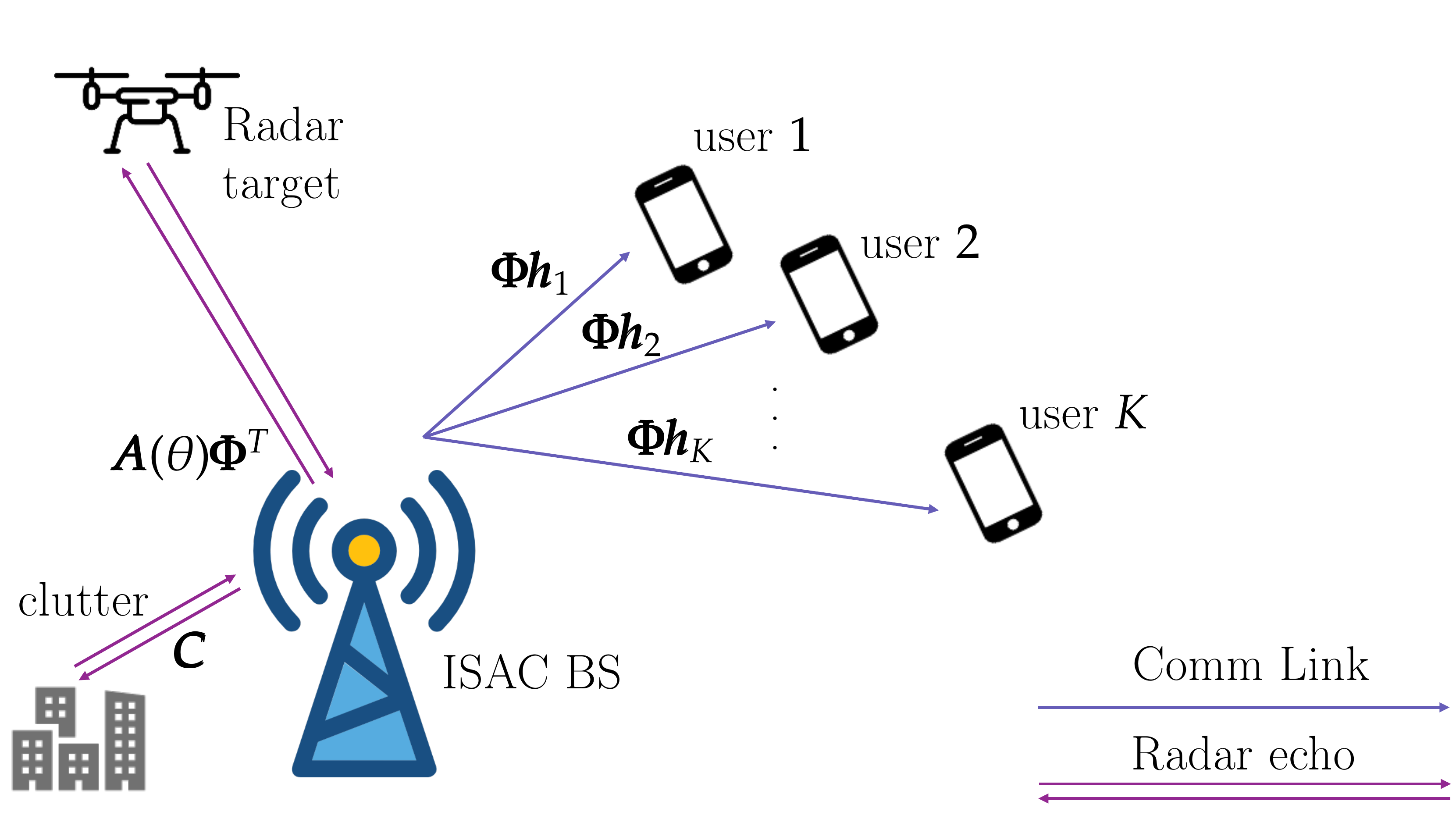}
\caption{
\textcolor{black}{\ac{ISAC}} scenario including an \textcolor{black}{\ac{ISAC} \ac{BS}}, an intended target with clutter and $K$ communication users.}
\label{fig:fig_1}
	\end{figure}
Consider an \ac{ISAC} system composed of $K$ single-antenna communication users, a radar target of interest, and a \textcolor{black}{\ac{ISAC}} \ac{BS}. Let the number of transmit and receive antennas at the \ac{BS} be $N_t$ and $N_r$, respectively. The communication users are considered to be located at arbitrary positions, whereas the target is supposed to be at a given angle $\theta_0$ from the \textcolor{black}{\ac{ISAC}} \ac{BS}. In the following, we describe the system model for both communication and radar sensing.
\subsection{Communication Model}
Consider the \textcolor{black}{\ac{ISAC}} \ac{BS} broadcasting a pilot signal in the \ac{DL} sense over its $N_t$ transmit antennas. 
In particular, let $\pmb{\Phi} \in \mathbb{C}^{L \times N_t}$ be the pilot matrix, to be designed, where $[\pmb{\Phi}]_{\ell,n}$ represents the $\ell^{th}$ pilot symbol transmitted over the $n^{th}$ antenna. In matrix notation, we have
\begin{equation}
	\pmb{\Phi}
	=
	\begin{bmatrix}
		\pmb{\Phi}_1 & \ldots & \pmb{\Phi}_L
	\end{bmatrix}^T.
\end{equation} 
Furthermore, let $\pmb{h}_k$ be the channel between \textcolor{black}{\ac{ISAC}} \ac{BS} and $k^{th}$ communication user. Then, the received signal in the \ac{DL} over \textcolor{black}{$L$} instances seen at the the $k^{th}$ communication user reads
\begin{equation}
	\label{eq:comm-model}
	\pmb{y}_k = \pmb{\Phi}\pmb{h}_k + \pmb{n}_k.
\end{equation}
\textcolor{black}{\input{Actions/stage-1-estimation}}In \eqref{eq:comm-model}, $\pmb{y}_k \in \mathbb{C}^{L \times 1}$ is the received signal vector at the $k^{th}$ user and $\pmb{n}_k$ is background noise, assumed to be white Gaussian \textcolor{black}{independently and identically distributed} with zero mean, 
\textcolor{black}{\input{Actions/C3o3_1.tex}}

We adopt the \ac{GMM} in order to model the communication channels between \textcolor{black}{\ac{ISAC}} \ac{BS} and the users.
A \ac{GMM} is a general model for multicasting, where each communication user is spatially distributed in a non-uniform geometric fashion. Even more, with a large number of mixtures, one can approximate any density with the aid of \ac{GMM} \cite{7736094,5089998}. 
 Typically, a \ac{GMM} describes an ensemble of Gaussian distributions, whereby increasing the number of Gaussian components can contribute to better describing the channel \cite{6819874}.
The \ac{PDF} of channel $\pmb{h}_m$ is modeled as \ac{GMM}, i.e. a weighted sum of Gaussian densities as follows,
\begin{equation}
	\label{eq:GMM}
	\phi_h(\pmb{h}_k) 
	=
	\sum\nolimits_{n = 1}^{N_k}
	\alpha_{k,n}
	\phi_n(\pmb{h }_k),
\end{equation}
\textcolor{black}{\input{Actions/C301_density.tex}} Furthermore, $\alpha_{k,n}$ are the so-called mixing coefficients, reflecting the probability that the $n^{th}$ Gaussian component, i.e. $\phi_n(\pmb{h}_k)$, is active in the  \eqref{eq:GMM}, therefore, we must have $\sum_{n = 1}^{N_k} \alpha_{k,n} = 1$, $\forall k$. Also, the number of Gaussian components modeling the $k^{th}$ communication user is $N_k$. Finally, as indicated in \cite{6819874}, the set $\lbrace \alpha_{k,n}, \pmb{\mu}_{k,n},\pmb{R}_{k,n} \rbrace_{n=1}^{N_k}$ are the \ac{GMM} parameters that model channel component $\pmb{h}_k$.

The communication users perform channel estimation, e.g. linear \ac{MMSE} estimation, \textcolor{black}{and then} feedback those estimates to the \textcolor{black}{\ac{ISAC}} \ac{BS}. The \textcolor{black}{\ac{ISAC}} \ac{BS} then uses those estimates to precode the modulated information as follows
\begin{equation}
	\label{eq:system-model-comm-for-stage-2}
	\pmb{Y}_c = \pmb{H} \pmb{W} \pmb{S}_c + \pmb{N}_c,	
\end{equation}
where $\pmb{H}  = \begin{bmatrix}
	\pmb{h}_1 & \ldots & \pmb{h}_K
\end{bmatrix}^T$ is the communication channel matrix over all users, assumed fixed within a certain block. Moreover, $\pmb{S}_c \in \mathbb{C}^{K \times B}$ belong to a certain constellation, e.g. $2^p-$\ac{QAM} and $B$ is the block length. In addition, $\pmb{W}$ is a communication precoding matrix (e.g. \ac{ZF} or \ac{MMSE} precoding) that utilizes the channel estimates reported by the users. \textcolor{black}{\input{Actions/stage-2-communications}}

\subsection{Sensing Model}
The radar sub-system within the \textcolor{black}{\ac{ISAC} \ac{BS}} system exploits the same pilot matrix as the one utilized for communication tasks, i.e. $\pmb{\Phi}$. In precise, the pilots transmitted over $N_t$ antennas and $L$ slots are designed to satisfy both dual sensing and communication functionalities. Assuming a colocated mono-static \ac{MIMO} radar setting, the received backscattered signal at the \textcolor{black}{\ac{ISAC}} \ac{BS} over $N_r$ receiving antennas can be written as \cite{4359542}
\begin{equation}
	\label{eq:yr}
	\pmb{Y}_r = 
	\gamma_0 \pmb{a}_{N_r}(\theta_0)\pmb{a}_{N_t}^T(\theta_0)\pmb{\Phi}^T +  
	\sum\limits_{i=1}^Q \gamma_i \pmb{a}_{N_r}(\theta_i)\pmb{a}_{N_t}^T(\theta_i)\pmb{\Phi}^T + 
	\pmb{Z}_r.
\end{equation}
\textcolor{black}{\input{Actions/C3o1_monostatic}}
The vectors $\pmb{a}_{N_r}(\theta) \in \mathbb{C}^{N_r \times 1}$ and $\pmb{a}_{N_t}(\theta) \in \mathbb{C}^{N_t \times 1}$ model the transmit/receive array steering vectors pointing towards angle $\theta$, respectively. For instance, if a \ac{ULA} array setting is adopted, then the steering vectors are given as follows
\begin{align}
	\label{eq:steering-1}
	\pmb{a}_{N_t}
	(\theta)
	&=
	\begin{bmatrix}
		1 & e^{j \frac{2\pi}{\lambda} d_t \sin(\theta)} & \ldots & e^{j \frac{2\pi}{\lambda} d_t(N_t-1) \sin(\theta)}
	\end{bmatrix}^T, \\
	\label{eq:steering-2}
	\pmb{a}_{N_r}
	(\theta)
	&=
	\begin{bmatrix}
		1 & e^{j \frac{2\pi}{\lambda} d_r \sin(\theta)} & \ldots & e^{j \frac{2\pi}{\lambda} d_r(N_r-1) \sin(\theta)}
	\end{bmatrix}^T,
\end{align}
where $\lambda$ is the wavelength and $d_t,d_r$ are the inter-element spacing between transmit and receive antennas, respectively. Since a co-located setting is utilized, the \ac{AoD} and \ac{AoA} of the backscattered component is the same.

We assume that the above reception, sampling and signal processing happen within the same time interval referred to as the \ac{CPI} \cite{10105893,8955956,mishra2019toward}, which is an interval where $\theta_0, \theta_1 \ldots \theta_Q$ and $\gamma_0,\gamma_1 \ldots \gamma_Q$ remain unchanged \cite{9540344}. The \ac{CPI} length depends on the mobility of objects in the channels and is typically a few milliseconds when objects move at speeds of tens of meters per second. 
On the other hand, the $\gamma_i$'s vary between different realizations of $\pmb{Y}_r$.\footnote{This occurs when the \textcolor{black}{\ac{ISAC}} \ac{BS} attempts to transmit the same pilot matrix $\pmb{\Phi}$ within another frame.} Therefore, following the Swerling-I model, we can assume that \textcolor{black}{$\gamma_i \sim \mathcal{CN}(0,\nu_i)$}\cite{6210398,4200703}. In addition, the $i^{th}$ clutter power is denoted as $\nu_i = \textcolor{black}{\mathbb{E}[ \vert \gamma_i \vert^2]}$.

Furthermore, the backscattered signal $\pmb{Y}_r \in \mathbb{C}^{N_r \times L}$ is the received radar vector and $\gamma_0$ is the complex channel gain of the reflected echo, containing two-way delay information between the \textcolor{black}{\ac{ISAC}} \ac{BS} and the radar target of interest. The angle $\theta$ represents the \ac{AoA} of the target relative to the \textcolor{black}{\ac{ISAC} \ac{BS}} receive array. 
Nevertheless, due to clutter presence, the $i^{th}$ clutter source with complex gain $\gamma_i$ is located at $\theta_i$ relative to the \textcolor{black}{\ac{ISAC}} \ac{BS}. In addition, $Q$ is the number of clutter components in the environment. The noise of the radar sub-system is i.i.d Gaussian modeled as $\pmb{Z}_r \sim \mathcal{N}(0,\sigma_r^2 \pmb{I}_N)$. In the next section, we introduce performance metrics to optimize the pilot matrix intended for both sensing and communication tasks.

\textcolor{black}{\input{Actions/GMM-radar-difference-justification}}

%% file: Actions/stage-1-estimation.tex
This is the first stage of a communication system, commonly referred to as the training phase, where the goal is to reliably estimation $\pmb{h}_1 \ldots \pmb{h}_K$. 
Note that no precoding is adopted and the goal here is to design $\pmb{\Phi}$ well-optimized for channel estimation, and target detection as will be described in Section \ref{sec:MI-for-sensing-metrics}.

%% file: Actions/C3o3_1.tex
i.e. $\boldsymbol{n}_k \sim$ $\mathcal{C N}\left(0, \sigma_k^2 \boldsymbol{I}_L\right)$.

%% file: Actions/C301_density.tex
where $\phi_n(\pmb{h}_k) = f_{\mathcal{CN}}(\pmb{h}_k \vert \pmb{\mu}_{k,n},\pmb{R}_{k,n})$, where $\pmb{\mu}_{k,n} \in \mathbb{C}^{N_t \times 1}$ and $\pmb{R}_{k,n} \in \mathbb{C}^{N_t \times N_t}$.

%% file: Actions/stage-2-communications.tex
Equation \ref{eq:system-model-comm-for-stage-2} represents the second stage after channel estimation, whereby $\pmb{W}$ is designed for precoding to maximize communication performance.
In this paper, we have adopted the \ac{ZF} precoding that makes use of the channel estimates given in the first stage. 

%% file: Actions/C3o1_monostatic.tex
The \acp{AoA} and \acp{AoD} are identical due to the monostatic sensing setting, i.e. the transmitter and receiver are assumed to very close, i.e. colocated. 
Indeed, due to geometric symmetry, the \ac{AoD}, which measures the angle of the path departing from the \textcolor{black}{\ac{ISAC}} \ac{BS} towards the target/clutter is equal to the angle of the echo joining the same target/clutter back to the \textcolor{black}{\ac{ISAC}} \ac{BS}.
This translates to identical \acp{AoA} and \acp{AoD} for both target and clutter.
It is worth noting that this is not the case for bistatic sensing, however.

%% file: Actions/GMM-radar-difference-justification.tex
In this work, we assumed a clear \ac{LoS} component between the target and the \ac{BS} for the radar channel, for e.g. a flying \ac{UAV}. This assumption aligns with scenarios where the \ac{BS} senses highly visible targets, as noted in the \ac{3GPP} feasibility study on \ac{ISAC} \cite{3gpp_tr22837_2023}, which includes use cases like \ac{UAV} trajectory tracking and detection. \ac{UAV} targets typically exhibit a strong \ac{LoS} channel, which differs significantly from the ground-based user communication channel. 
For the communication channel, the scenario is more complex due to effects like shadowing, multipath, and \ac{NLoS} conditions. These factors often lead to distributions such as Rician (via $K$-factor modeling) or Nakagami for severe fading. To address this variability, we adopted a \ac{GMM} model, which is versatile for approximating unknown channel distributions. Even when the actual channel is not inherently a \ac{GMM}, its flexibility allows it to approximate a wide range of distributions accurately \cite{4115301}.
Moreover, \ac{GMM} models are particularly useful under the weakest assumptions about the channel, as they can represent any continuous probability distribution with arbitrary precision  \cite{nguyen2020approximation}. This makes them a robust choice for generalizing \ac{ISAC} pilot design in uncertain communication channel conditions.

%% file: sections/performance-metrics.tex
Radar waveforms are designed to enhance various sensing capabilities, such as detection, imaging and location estimation accuracy. On the other hand, \textcolor{black}{communication waveforms maximize the amount of information reliably transferred between} the \ac{BS} and the communication users. In this study, the aim is to design a common pilot matrix to simultaneously optimize the sensing detection potential at the \textcolor{black}{\ac{ISAC}} \ac{BS} and channel estimation reliability at all communication users.
In this section, we attempt to unify the communication and sensing performance metrics via appropriate mutual information based indicators. 

\subsection{Mutual Information for Communications}
The \ac{MI} is a special instance of the more general quantity, that is the relative entropy, which measures the distance between probability distributions \cite{cover1999elements}.
Following \cite{cover1999elements}, the \ac{MI} between the received \ac{DL} signal at the $k^{th}$ communication user and the $k^{th}$ channel vector can be expressed as
\begin{equation}
	I(\pmb{y}_k; \pmb{h}_k)
	=
	h(\pmb{y}_k)
	-
	h(\pmb{y}_k \vert \pmb{h}_k ),
\end{equation}
where $h(\pmb{y}_k)$ is the differential entropy of the received signal $\pmb{y}_k$ and $h(\pmb{y}_k \vert \pmb{h}_k )$ stands for differential entropy of $h(\pmb{y}_k)$ given $\pmb{h}_k$. In the following, we propose an \ac{MI}-based metric associated with the $k^{th}$ communication user. \\\\
\textbf{Proposition 1}: \textit{For a given pilot matrix $\pmb{\Phi}$, the \ac{MI} between $\pmb{y}_k$ and $\pmb{h}_k$ can be approximated as $I(\pmb{y}_k; \pmb{h}_k) \simeq  \mathcal{M}_k^{\tt{comm}}(\pmb{\Phi})$ where}
\begin{equation}
\label{eq:M-comm-k}
	\mathcal{M}_k^{\tt{comm}}(\pmb{\Phi}) \triangleq
	- \log \Big(\sum\nolimits_{n=1}^{N_k}  \frac{\alpha_{k,n} e^{-\beta_{k,n}(\pmb{\Phi})}}{\det (\pmb{\Sigma}_{k,n}(\pmb{\Phi})) } \Big)  +
	\tt{cnst},
\end{equation} 
\textit{and}
\begin{align}
	\label{beta-expression}
	\beta_{k,n}(\pmb{\Phi}) &= \bar{\pmb{\mu}}_{k,n}^H\pmb{\Phi}^H \pmb{\Sigma}_{k,n}^{-1}(\pmb{\Phi})\pmb{\Phi}\bar{\pmb{\mu}}_{k,n}, \\
	\label{eq:mu_bar}
	\bar{\pmb{\mu}}_{k,n} &=\sum\nolimits_{n' = 1}^{N_k}\alpha_{k,n'}\pmb{\mu}_{k,n'} - \pmb{\mu}_{k,n}, \\
	\label{eq:Sigma_Phi}
	\pmb{\Sigma}_{k,n}(\pmb{\Phi}) &=  \pmb{\Phi}\pmb{R}_{k,n}\pmb{\Phi}^H + \sigma_k^2 \pmb{I}_L.
\end{align}
\textit{Furthermore, ${\tt{cnst}}$ is a term independent of $\pmb{\Phi}$.}

\textbf{Proof}: See \textbf{Appendix \ref{app:MI-comm}}.

\subsection{Mutual Information for Sensing}
\label{sec:MI-for-sensing-metrics}
We consider that the \textcolor{black}{\ac{ISAC}} \ac{BS} listens to backscattering reflections and observes the signal based on two hypothesis. The first being $\mathcal{H}_0$, where it assumes the absence of any target, therefore observing interference and background noise. On the other hand, the alternative hypothesis, i.e. $\mathcal{H}_1$, where the signal is present and is superimposed on top of interference and background noise. To this end, the \textcolor{black}{\ac{ISAC} \ac{BS}} listens over $L$ samples and considers the following hypothesis testing problem,
\begin{equation}
\label{eq:Hypothesis-test-1}
	\begin{cases}
		\mathcal{H}_0: \qquad \pmb{y}_r =  \pmb{c} + \pmb{n}_r, \\ 
		\mathcal{H}_1: \qquad \pmb{y}_r = \pmb{d} +\pmb{c} + \pmb{n}_r.
	\end{cases}
\end{equation}
For convenience, we have directly expressed the hypothesis testing problem using vectorized quantities. In particular, $\pmb{y}_r = \ve(\pmb{Y}_r) \in \mathbb{C}^{N_rL \times 1}$, and the signal of interest, i.e. $\pmb{d}$, and the clutter vector $\pmb{c}$ can be expressed as 
%\begin{align}
%		\pmb{d} 
%		&=
%		\gamma_0
%		\begin{bmatrix}
%		\pmb{a}_{N_r}(\theta_0)\pmb{a}_{N_t}^T(\theta_0) \pmb{\Phi}_1 \\
%		\pmb{a}_{N_r}(\theta_0)\pmb{a}_{N_t}^T(\theta_0) \pmb{\Phi}_2 \\
%		\vdots \\	
%		\pmb{a}_{N_r}(\theta_0)\pmb{a}_{N_t}^T(\theta_0) \pmb{\Phi}_L \\
%	\end{bmatrix} 		\triangleq
%		\gamma_0 \pmb{\mu}_0,	\\
%		\pmb{c} 
%		&=
%		\sum_{i=1}^Q
%		\gamma_i
%		\begin{bmatrix}
%		\pmb{a}_{N_r}(\theta_i)\pmb{a}_{N_t}^T(\theta_i) \pmb{\Phi}_1 \\
%		\pmb{a}_{N_r}(\theta_i)\pmb{a}_{N_t}^T(\theta_i) \pmb{\Phi}_2 \\
%		\vdots \\	
%		\pmb{a}_{N_r}(\theta_i)\pmb{a}_{N_t}^T(\theta_i) \pmb{\Phi}_L \\
%	\end{bmatrix}\triangleq
%		\sum\limits_{i=1}^Q \gamma_i \pmb{\mu}_i.	
%\end{align}
\begin{align}
		\pmb{d} 
		&=
		\gamma_0
		\ve(\pmb{a}_{N_r}(\theta_0)\pmb{a}_{N_t}^T(\theta_0)\pmb{\Phi}^T )		\triangleq
		\gamma_0 \pmb{\mu}_0,	\\
		\label{eq:clutter-vec-expression}
		\pmb{c} 
		&=
		\sum\nolimits_{i=1}^Q
		\gamma_i
		\ve(\pmb{a}_{N_r}(\theta_i)\pmb{a}_{N_t}^T(\theta_i)\pmb{\Phi}^T )\triangleq
		\sum\nolimits_{i=1}^Q \gamma_i \pmb{\mu}_i.	
\end{align}
\textcolor{black}{\input{Actions/multi-target-case}}
An equivalent, and more convenient, way of expressing the hypothesis test in \eqref{eq:Hypothesis-test-1} is 
\begin{equation}
\label{eq:Hypothesis-test-2}
	\begin{cases}
		\mathcal{H}_0: \ \pmb{z} \sim \mathcal{CN}(\pmb{0},\pmb{I}), \\ 
		\mathcal{H}_1: \ \pmb{z} \sim \mathcal{CN}(\pmb{0},(\pmb{R}_{\pmb{cc}} + \sigma_r^2 \pmb{I})^{-\frac{1}{2}}\pmb{R}_{\pmb{dd}}(\pmb{R}_{\pmb{cc}} + \sigma_r^2 \pmb{I})^{-\frac{1}{2}} + \pmb{I}),
	\end{cases}
\end{equation}
where $\pmb{R}_{\pmb{cc}}$ and $\pmb{R}_{\pmb{dd}}$ are the covariance matrices of $\pmb{c}$ and $\pmb{d}$, respectively. Also, $\pmb{z} = (\pmb{R}_{\pmb{cc}} + \sigma_r^2 \pmb{I})^{-\frac{1}{2}}\pmb{y}_r$. In general, the larger the \ac{MI} between the parameters of interest and the observed data, the more information we have about the target of interest \cite{259642}. Based on this, we have the following \ac{MI}-based performance metric for radar sensing, \\
\textbf{Proposition 2}: \textit{For a given pilot matrix $\pmb{\Phi}$ and large enough number of receive antennas, i.e. $N_r$, the \ac{MI} related to the hypothesis testing problem in \eqref{eq:Hypothesis-test-1} can be approximated as $I(\pmb{Y}_r;\pmb{\Theta},\pmb{\nu} \vert \pmb{\Phi}) \simeq \mathcal{M}^{\tt{sense}}(\pmb{\Phi})$, where}
\textcolor{black}{
\begin{equation}
\label{eq:M-sense}
	\mathcal{M}^{\tt{sense}}(\pmb{\Phi}) \triangleq \log \left(1+ \frac{\nu_0}{\sigma_r^2 } \Vert \pmb{\mu}_0 \Vert^2 -\frac{\nu_0}{\sigma_r^2 }\sum_{i=1}^Q  \frac{  \nu_i\vert\pmb{\mu}_0^H\pmb{\mu}_i \vert^2}{\sigma_r^2 + \nu_i \Vert \pmb{\mu}_i \Vert^2}\right),
\end{equation}
}
\textcolor{black}{\input{Actions/C3o1_definethetanu}}
\textbf{Proof}: See \textbf{Appendix \ref{app:MI-sense}}.  \\
The \ac{MI}-based performance metric in \textbf{Proposition 2} provides various insights. First, the second term appearing in the $\log$ is the contribution in the look-direction towards a target of interest located at direction $\theta_0$. The third term appearing within the $\log$ is due to the presence of the $Q$ clutter components. Therefore, the maximization of the proposed \ac{MI}-based metric, i.e. $\mathcal{M}^{\tt{sense}}(\pmb{\Phi})$, aims at maximizing the difference between the power found in the look direction and that of the clutter components, in $\log$ sense. In the next sub-section, we aim at integrating both \ac{MI}-based metrics for optimizing communication and sensing performance.

%% file: Actions/multi-target-case.tex
In practice, multiple sensing target may co-exist. The detection problem herein can be extended to multiple targets by scanning multiple directions in a time division manner.
During the training phase for communications, one can optimize the pilots for a set of direction(s), which are to be transmitted to scan the environment.

%% file: Actions/C3o1_definethetanu.tex
\textit{where $\pmb{\Theta} = \begin{bmatrix}
	\theta_0 \ldots \theta_Q
\end{bmatrix}$ and $\pmb{\nu} = \begin{bmatrix}
	\nu_0 \ldots \nu_Q
\end{bmatrix}$.}

 

%% file: sections/optimization-problems.tex
In this section, we describe an optimization framework enabling us to generate orthogonal pilots that optimize the mutual information for communication and sensing.
Since multiple \ac{MI}-based objectives are to be jointly optimized, an appropriate \ac{MOOP} \cite{6924852} can be constructed. To this end, in order to jointly optimize both communication and sensing, we consider the following problem
\textcolor{black}{
\begin{equation}
	\label{eq:MOOP1}
\input{equations/moop1.tex}
\end{equation}}
\textcolor{black}{\input{Actions/powerbudgetdefinition}}\textcolor{black}{\input{Actions/why_assumption_L_l_Nt}}

\textcolor{black}{\input{Actions/remark_on_optimality.tex}}

In equation \eqref{eq:MOOP1}, the \ac{MOOP} aims at maximizing all sensing and multi-communication \ac{MI}s simultaneously. This type of optimization problem appears under different names in the literature, such as a vector optimization \cite{boyd2004convex} or multi-criteria optimization \cite{5336784}.
Notice that an orthogonality constraint is enforced on the pilot matrix through $\pmb{\Phi}\pmb{\Phi}^H = \pmb{I}$ because orthogonal pilot patterns are commonly employed and desired for multi-channel estimation. 
Generally speaking, orthogonal pilot sequences represent one approach to eliminate the inter-cell interference.
In addition, least squares channel estimators become straightforward to implement as no matrix inversion of $\pmb{\Phi}$ will be required. 

\textcolor{black}{\input{Actions/challenges.tex}}

Now, since the objective function in \eqref{eq:MOOP1} is multi-dimensional, we can transform it, utilizing a technique referred to as \textit{scalarization}, with the aid of a goal function \cite{6924852}, hereby denoted as $\psi:\mathbb{R}^{K+1} \rightarrow \mathbb{R}$. More precisely, this goal function describes an \ac{ISAC} tradeoff between the different \ac{MI}-based sensing and communication objectives. There are a number of goal function options to be aware of, the simplest being the weighted-sum function, i.e. $\psi(\pmb{x}) = \sum_k w_k \pmb{x}_k$. Another choice includes the geometric mean, which is useful when the underlying objectives have different numerical ranges. Since our objectives are all \ac{MI}-based, hence are of the same nature, then the geometric mean function does not seem to be useful. Other choices are the weighted Chebyshev function and the distance goal function. 

Adopting the scalarization technique via weighted-sum, the scalarized version of the \ac{MOOP} in \eqref{eq:MOOP1} can be written as 
\begin{equation}
	\label{eq:scalarization1}
\begin{aligned}
(\mathcal{P}_{\tt{s}}):
\begin{cases}
\max\limits_{\pmb{\Phi} }&  \mathcal{M}^{\tt{ISAC}}(\pmb{\Phi}) \triangleq \rho \mathcal{M}^{\tt{comm}}(\pmb{\Phi}) + (1-\rho)\mathcal{M}^{\tt{sense}}(\pmb{\Phi}) \\
\textrm{s.t.}
 & \pmb{\Phi}\pmb{\Phi}^H = \pmb{I}.
\end{cases}
\end{aligned}
\end{equation}
where $\mathcal{M}^{\tt{comm}}(\pmb{\Phi}) = \sum_k w_k \mathcal{M}_k^{\tt{comm}}(\pmb{\Phi})$ and $\sum_k w_k = 1$. In the above formulation, the design parameter $\rho$ balances the \ac{ISAC} tradeoff between sensing, in terms of detection performance seen through $\mathcal{M}^{\tt{sense}}(\pmb{\Phi})$ and the communication channel estimation performance, which is seen through $\mathcal{M}^{\tt{comm}}(\pmb{\Phi})$. In other words, increasing $\rho$ gives a higher priority onto the communication sub-system and decreasing it prioritizes sensing. For communications, the $k^{th}$ user is given a preference associated with value $w_k$. For equal performance along communication users, one can set \textcolor{black}{$w_1 = \cdots = w_K = \frac{1}{K}$}. Observing problem $(\mathcal{P}_{\tt{s}})$ in \eqref{eq:scalarization1}, the cost $ \mathcal{M}^{\tt{ISAC}}(\pmb{\Phi})$ in conjunction with the orthogonality constraint leave us with a highly non-convex and non-linear optimization problem. 
In the next section, we propose an algorithm to directly solve $(\mathcal{P}_{\tt{s}})$.

%% file: equations/moop1.tex
\begin{aligned}
(\mathcal{P}_{\tt{MOOP}}):
\begin{cases}
\max\limits_{\pmb{\Phi} }&  \begin{bmatrix}
	\mathcal{M}_1^{\tt{comm}}(\pmb{\Phi}),
	 \ldots,
	 \mathcal{M}_K^{\tt{comm}}(\pmb{\Phi}), 
	 \mathcal{M}^{\tt{sense}}(\pmb{\Phi})
\end{bmatrix} \\
\textrm{s.t.}
 & \pmb{\Phi}\pmb{\Phi}^H = \frac{P}{L} \pmb{I},
\end{cases}
\end{aligned}

%% file: Actions/powerbudgetdefinition.tex
where $P$ is a given power budget. 
An increase of power budget can indeed translate to an increase in sensing and communication mutual information, as can be inferred from equations \eqref{eq:M-comm-k} and \eqref{eq:M-sense}. 
In this paper, we fix the power budget to $\frac{P}{L}=1$ and aim at optimizing the pilot matrix itself that maximizes both sensing and communication mutual information, without simply increasing power.

%% file: Actions/why_assumption_L_l_Nt.tex
We note that an assumption of $L < N_t$ is required in order to guarantee an orthonormal solution of the problem in \eqref{eq:scalarization1}.
In the system defined by the constraint $\boldsymbol{\Phi} \boldsymbol{\Phi}^H = \boldsymbol{I}$, there are $L N_t$ unknowns and $L^2$ equations. Setting $L < N_t$ provides more degrees of freedom than constraints, enabling the optimization of the \ac{MI} objective in \eqref{eq:scalarization1} as described in the updated manuscript. Conversely, if $L \geq N_t$, the condition $\boldsymbol{\Phi} \boldsymbol{\Phi}^H = \boldsymbol{I}$ cannot be satisfied because the rank of $\boldsymbol{\Phi} \boldsymbol{\Phi}^H$ would be limited to $N_t$.

%% file: Actions/remark_on_optimality.tex
It is crucial to define an order relation to properly define our \ac{MOOP} in $(\mathcal{P}_{\tt{MOOP}})$. For this, we have\\
\textbf{Remark}: For any two vectors $\pmb{a},\pmb{b} \in \mathbb{R}^{K+1}$ and $\pmb{a} \neq \pmb{b}$, we say that $\pmb{b}$ \textit{Pareto dominates} $\pmb{a}$ if and only if $\pmb{a}_i \leq \pmb{b}_i$ for all $i = 1\ldots K+1$.
This ordering can be denoted as $\pmb{a} \leqslant \pmb{b}$.
A "stricter" order can be defined by enforcing a strict inequality component-wise.
In both cases, $\leqslant$ defines a strict partial order on the multidimensional Euclidean space $\mathbb{R}^{K+1}$.
This order relation is commonly referred to as \textit{Pareto order} or \textit{component-wise order} \cite{ehrgott2005multicriteria}.

%% file: Actions/challenges.tex
Before we proceed, taking a closer look at problem \eqref{eq:MOOP1} suggests that the fundamental challenge of directly optimizing \eqref{eq:MOOP1} poses a hurdle due to conflicting objectives.
In other words, the optimal solution of a given function is $\mathcal{M}_1^{\tt{comm}}(\pmb{\Phi})$ is not necessarily that of $\mathcal{M}^{\tt{sense}}(\pmb{\Phi})$, and improving one can indeed worsen the other. 
In turn, this means that we may not have a global optimum because there are only subjectively optimal solutions \cite{6924852}.
Technically speaking, the so-called \textit{utopia point}, i.e. the point that maximizes each individual mutual information objective, does not necessarily lie in the attainable objective set \cite{6924852}. Therefore, the problem is nontrivial and may not have a global optimum. 
It should be noted that, addressing problem \eqref{eq:MOOP1} as-is poses technical challenges, it is theoretically possible. 
A straightforward way would be to sample many feasible pilot matrices satisfying the orthogonality constraint $\boldsymbol{\Phi} \boldsymbol{\Phi}^H=\boldsymbol{I}$.
Besides the heavy complexity involved, this approach does not ensure that any of the sampled pilot matrices lies exactly on the Pareto boundary.
Another approach that can be mentioned within this context is the so-called bisection method for Pareto boundary exploration.
In short, the bisection method \cite{6924852} aims to search in given directions and adjust the magnitude of the search vector (through bisection), in order to hit the Pareto frontier. 
Two challenges can be directly enumerated: $(i)$ The search direction is not-known in advance as the goal is to find the entire set of Pareto points; and $(ii)$ even if we were given the optimal direction, the bisection method involves a membership test, which is itself a complicated problem, as it involves finding the optimal matrix for membership at each iteration. 
For these reasons, we resort to a linearization method.

%% file: sections/optimization01.tex
\subsection{Projected Gradient Descent over the Stiefel Manifold}
\begin{figure}[t]
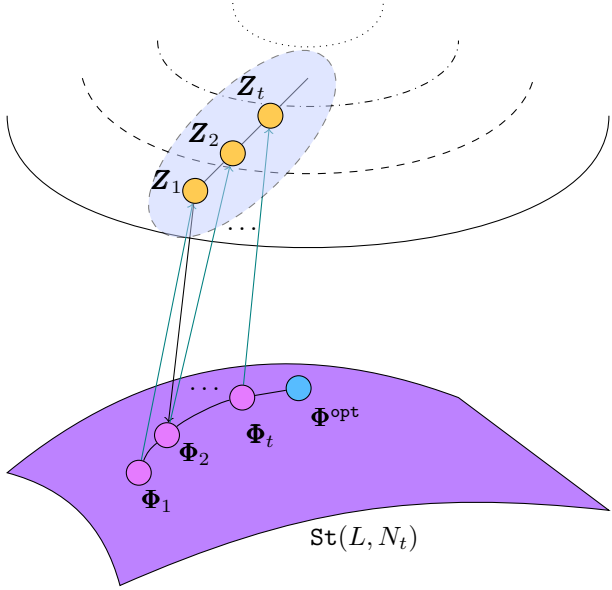

	\centering
	\ctikzfig{./figures/PGD}
	\caption{An illustration depicting the convergence of the projected gradient descent method for \ac{ISAC} pilot-design. The point $\pmb{\Phi}^{\tt{opt}}$ is the optimal design matrix, which \textcolor{black}{locally} maximizes $\mathcal{M}^{\tt{ISAC}}(\pmb{\Phi})$ on the Stiefel manifold. }
	\label{fig:PGD}
\end{figure}

We design a projected gradient descent method to optimize problem $(\mathcal{P}_{\tt{s}})$. In other words, the projected gradient descent iterates as follows
\begin{align}
	\label{eq:GD-step}
	\pmb{Z}_{t+1} 
	&=
	\pmb{\Phi}_{t}
	+
	\gamma
	\nabla_{\pmb{\Phi}}
	\mathcal{M}^{\tt{ISAC}}(\pmb{\Phi})
	\Big\vert _{\pmb{\Phi} = \pmb{\Phi}_{t}}, \\
	\label{eq:project-step}
	\pmb{\Phi}_{t+1}
	&=
	\pmb{\pi}(\pmb{Z}_{t+1} ),
\end{align}
where 
\begin{equation}
	\label{eq:GD-project}
	\pmb{\pi}(\pmb{Y})
	=
	\argmin_{\pmb{X} \in {\tt{St}}(L,N_t)}
	\Vert
	\pmb{X}
	-
	\pmb{Y}
	\Vert^2_F.
\end{equation}
where ${\tt{St}}(N_t,L) = \lbrace \pmb{X} \in \mathbb{C}^{L 
\times N_t}: \pmb{X}\pmb{X}^H = \pmb{I}_{L}\rbrace$ is the Stiefel manifold, which "boils down" to the unit-sphere manifold for $L = 1$. Note that for the case of $N_t > L$, $\pmb{\pi}(\pmb{Y}) = \pmb{U}\pmb{I}_{L,N_t}\pmb{V}^H$, where $\pmb{I}_{L,N_t} \in \mathbb{R}^{L \times N_t}$ is a rectangular diagonal matrix with all of its diagonal entries set to $1$. Also $\pmb{Y} = \pmb{U}\pmb{\Lambda}\pmb{V}^H$ represents the \textcolor{black}{singular value decomposition} operation. Observe that \eqref{eq:GD-step} is a classical gradient descent step and \eqref{eq:project-step} projects the current intermediate point onto the closest point within the feasible set, i.e. the Stiefel manifold, via $\pmb{\pi}(\pmb{Y})$. In this way, any iteration of the projected gradient algorithm is guaranteed to output an orthogonal pilot, while maximizing the cost $\mathcal{M}^{\tt{ISAC}}(\pmb{\Phi})$.

We now shed light on the gradients, by noting that their expressions can be computed in closed-form. Using the expression defined of $\mathcal{M}^{\tt{ISAC}}(\pmb{\Phi})$ defined in \eqref{eq:scalarization1}, and with the help of \eqref{eq:M-comm-k} and \eqref{eq:M-sense}, we can write
\begin{equation}
	\nabla_{\pmb{\Phi}}
	\mathcal{M}^{\tt{ISAC}}(\pmb{\Phi})
	=
	\rho \sum_k w_k \nabla_{\pmb{\Phi}}\mathcal{M}_k^{\tt{comm}}(\pmb{\Phi}) + (1-\rho)\nabla_{\pmb{\Phi}}\mathcal{M}^{\tt{sense}}(\pmb{\Phi}),
\end{equation}
where the expression of $\nabla_{\pmb{\Phi}}\mathcal{M}_k^{\tt{comm}}(\pmb{\Phi})$ is given in \textbf{Appendix \ref{app:dMI-comm}} and the expression of $\nabla_{\pmb{\Phi}}\mathcal{M}^{\tt{sense}}(\pmb{\Phi})$ is given in \textbf{Appendix \ref{app:dMI-sense}}. At this point, the projected gradient descent method is fully defined and can be executed with the defined gradients.
We now analyze the convergence properties of the iterative algorithm.

\subsection{Convergence Properties}
Unlike classical analysis, where the cost and the projection set are assumed to be convex, this analysis studies the convergence properties of the projected gradient descent of the non-convex pilot design problem. 
Before we proceed, we make some assumptions regarding the structure of the function $\mathcal{M}^{\tt{ISAC}}(\pmb{\Phi})$. The following definitions provide lower/upper bounds on specific classes of differentiable, and possibly non-convex, functions\textcolor{black}{.} \\
\textcolor{black}{\textbf{Definition 1 \mypcite{10.1093/imaiai/iay002}:}}  A function $f:\mathbb{R}^n \rightarrow \mathbb{R}$ is said to be $(\alpha,\beta,\epsilon)-$\ac{RSS} if it is continuously differentiable over a possibly non-convex region $\mathcal{S} \subseteq \mathbb{R}^n$ and for every $\pmb{x}_1,\pmb{x}_2 \in \mathcal{S} $, we have that 
\begin{equation}
\label{eq:RSS}
	f(\pmb{x}_2) - f(\pmb{x}_1) - \nabla f^T(\pmb{x}_1)(\pmb{x}_2-\pmb{x}_1) \leq \frac{\beta}{2} \Vert \pmb{x}_1 - \pmb{x}_2 \Vert^2 + \frac{\alpha}{2} \epsilon^2.
\end{equation} 
\textcolor{black}{\textbf{Definition 2 \mypcite{10.1093/imaiai/iay002}:}}  A function $f:\mathbb{R}^n \rightarrow \mathbb{R}$ is said to be $(\alpha,\epsilon)-$\ac{RSC} if it is continuously differentiable over a possibly non-convex region $\mathcal{S} \subseteq \mathbb{R}^n$ and for every $\pmb{x}_1,\pmb{x}_2 \in \mathcal{S} $, we have that 
\begin{equation}
\label{eq:RSC}
	f(\pmb{x}_2) - f(\pmb{x}_1) - \nabla f^T(\pmb{x}_1)(\pmb{x}_2-\pmb{x}_1) \geq \frac{\alpha}{2} \Vert \pmb{x}_1 - \pmb{x}_2 \Vert^2 - \frac{\alpha}{2} \epsilon^2.
\end{equation}
Given the \ac{RSC} and \ac{RSS} of a continuously differentiable function, the term $\epsilon$ captures the non-convexity of $f$ along $\mathcal{S} \subseteq \mathbb{R}^n$. Indeed, with $\epsilon = 0$, one can see that the function is strongly convex and smooth. Hence, $\epsilon,\alpha,\beta$ can be seen as parameters that trade-off smoothness and convexity.
Note that $(\alpha,\epsilon)-$\ac{RSC} assumptions were also used to establish theoretical results to quantify local optima of regularized estimators, where loss/penalty function are allowed to be non-convex over its associated domain \cite{loh2013regularized}.

Provided these definitions, we have the following convergence result on the pilot design projected gradient descent method. \\
\textbf{Theorem 1}: \textit{For any random initialization of the pilot matrix $\pmb{\Phi}_0$ and given $\alpha,\beta,\epsilon$ such that function $-\mathcal{M}^{\tt{ISAC}}(\pmb{\Phi})$ is $(\alpha,\epsilon)-$\ac{RSC} and $(\alpha,\beta,\epsilon)-$\ac{RSS}, the projected gradient descent \ac{ISAC}-based pilot design algorithm described in \eqref{eq:GD-step}, \eqref{eq:project-step}, \eqref{eq:GD-project} converges to $\pmb{\Phi}_{\infty}$, such that}
\begin{equation}
\label{eq:convergence-bound}
	\mathcal{M}^{\tt{ISAC}}(\pmb{\Phi}^{{\tt{opt}}}) 
	-
	\mathcal{M}^{\tt{ISAC}}(\pmb{\Phi}_{\infty})
	\leq
	\frac{\alpha + \frac{\beta}{2}}{1-\frac{\beta}{\alpha}}
	\epsilon^2,
\end{equation}
\textit{where $\pmb{\Phi}^{{\tt{opt}}}$ is the \textcolor{black}{local} maximum of $\mathcal{M}^{\tt{ISAC}}(\pmb{\Phi})$.}

\textbf{Proof}: See \textbf{Appendix \ref{app:convergence-analysis}}.

Therefore, \textbf{Theorem 1} reveals that the projected gradient descent method converges to the true maximizer $\pmb{\Phi}^{{\tt{opt}}}$, up to a certain precision level controlled by $\epsilon$. An illustration showing the points related to the projected gradient descent method is given in Fig. \ref{fig:PGD}. The pilots generated by the method are projected back onto the Steifel manifold ${\tt{St}}(L,N_t)$ to give $\pmb{\Phi}_t$ whenever a new descent update is available, i.e. $\pmb{Z}_{t-1}$.

\textcolor{black}{\input{Actions/benefits.tex}}
\begin{figure}[t]
	\centering
	\includegraphics[width=0.9\linewidth]{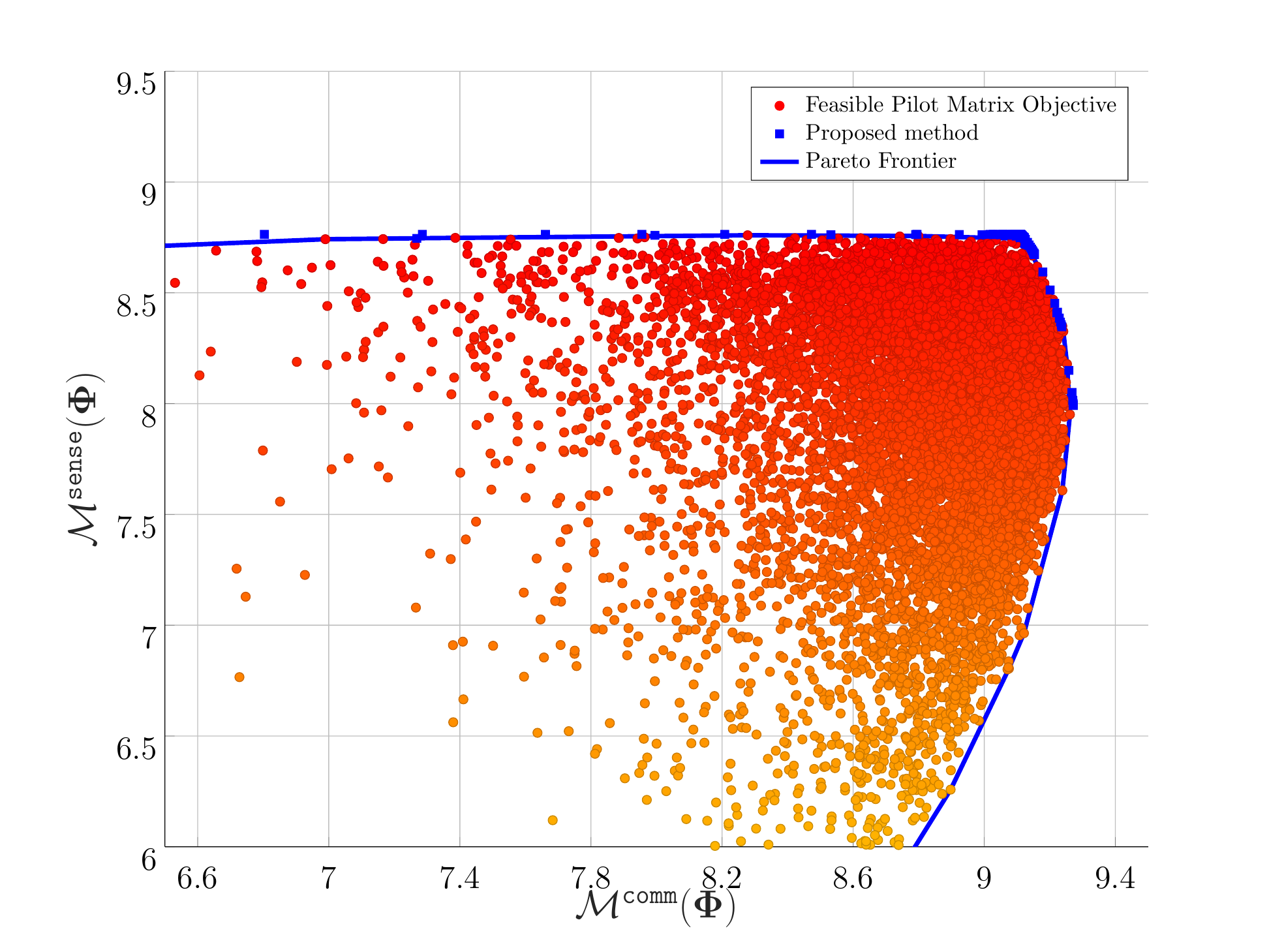}
	\caption{\textcolor{black}{The feasible pilot matrices region with the corresponding Pareto Frontier and the proposed method that samples on the frontier for different values of $\rho$. The single-user case.}}
	\label{fig:paretofront}
\end{figure}

%% file: Actions/benefits.tex
To illustrate the benefits of the proposed algorithm, we take a simple single-user example depicted in Fig. \ref{fig:paretofront}.
This example is intended to show that the proposed projected gradient descent method can efficiently sample the Pareto frontier over different values of sensing and communication balancing parameter $\rho$. 
To generate the feasible pilot matrix region, we exhaustively sample over $10000$ orthogonal pilot matrices and compute the sensing and communication mutual informations. 
We then compute the Pareto frontier, followed by launching the proposed method for random initializations for different values of $\rho$.
This simulation tells us that we can almost certainly achieve the Pareto frontier through simple recursion operations dictated in equations \eqref{eq:GD-step}, \eqref{eq:project-step} and \eqref{eq:GD-project}, without having to resort to more complicated multi-objective methods.

%% file: sections/motivation.tex
\subsection{MI in the context of communication channel estimation}
\label{subsec:MI-for-comm}
\input{Actions/motivation-MI-for-comm}

\begin{figure}[t]
	\centering
	\includegraphics[width=0.7\linewidth]{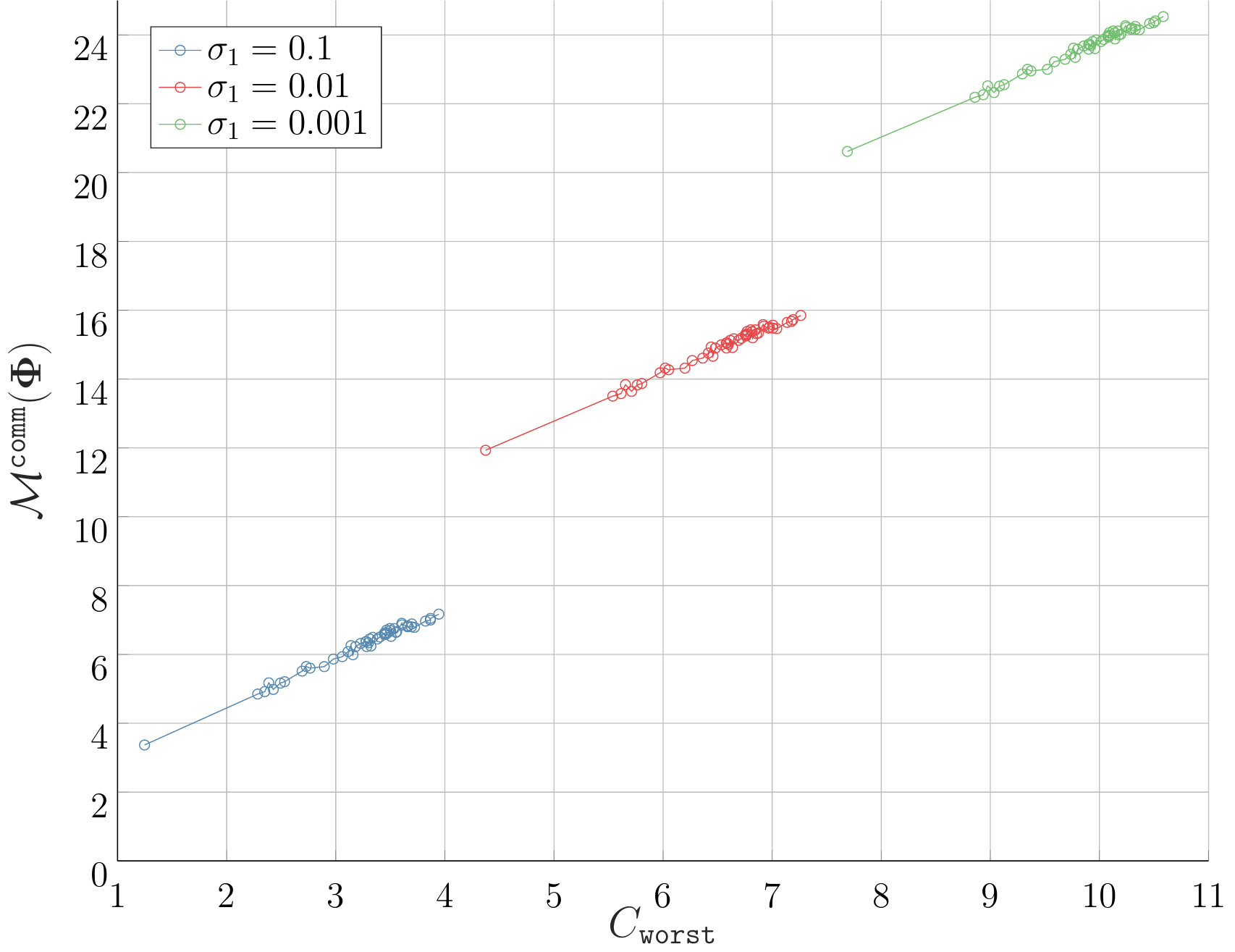}
	\caption{\textcolor{black}{The proposed \ac{MI}-based communication metric as a function of worst case channel capacity. Case of single-user and different noise levels.}}
	\label{fig:MIcomm_vs_Cworst}
\end{figure}

\subsection{MI in the context of sensing}
\label{subsec:MI-for-sensing}
\input{Actions/motivation-for-sensing}

\subsection{MI as a unifying \ac{ISAC} metric}
\label{subsec:MI-for-ISAC}
\input{Actions/motivation-unify}

%% file: Actions/motivation-MI-for-comm.tex
Many natural questions arise at this point. 
A very fundamental question to ask is the following: 
\textit{"Why would we adopt an \ac{MI}-based metric for channel estimation ?"}
Another follow-up question can be the following: 
\textit{"How does the adopted \ac{MI}-based metric relate to the achievable rate ?"}
At first sight, the adopted metric may seem somewhat absurd, however, we show that the adopted \ac{MI} metric, namely $\mathcal{M}^{\tt{comm}}(\pmb{\Phi})$, directly impacts the achievable rate.

\input{Actions/motivation-MI-for-comm-part1}
\input{Actions/motivation-MI-for-comm-part2}
\input{Actions/motivation-MI-for-comm-part3}

%% file: Actions/motivation-MI-for-comm-part1.tex
In \cite{1193803}, the authors study the impact of channel estimation errors, taking into account the training time, power allocation strategies between data transmission and training sequence, and the impact these factors have on the capacity of a channel given an estimate of the channel as prior information. 
In this regard, two stages can be naturally distinguished: 
$(i)$ the training stage utilizing $L$ time instances where a pilot sequence $\pmb{\Phi}$ is passed through the channel $\pmb{H}$, where the task here is to estimate the channel, and provide this channel estimation $\widehat{\pmb{H}}$ by observing $\pmb{Y}_t = \pmb{\Phi}\pmb{H} + \pmb{N}_t.$, where $\pmb{Y}_t $ and $\pmb{N}_t$ are concatenated versions of $\pmb{y}_k$ and $\pmb{n}_k$ given in equation \eqref{eq:comm-model}. The channel estimate is then passed to the subsequent stage where,
$(ii)$ the data transmission stage utilizing $B$ data blocks, which uses $\widehat{\pmb{H}}$ as prior information to decode the transmitted data.
Given this, the lower bound (which is tight under certain conditions \cite{1193803}) on the channel capacity is \cite{1193803}
\textcolor{black}{
\begin{equation}
\label{eq:Cworse}
	C_{\text {worst }}
	=
	\mathbb{E} \left[  \frac{B}{B+L} \log \operatorname{det}\left(\pmb{I}
	+\frac{\sigma_{\widehat{\pmb{H}}}^2}{1+ \sigma_{\widetilde{\pmb{H}}}^2} \frac{\bar{\pmb{H}}^* \bar{\pmb{H}}^T}{N_t}\right) \right],
\end{equation}
}
where $\bar{\pmb{H}}$ is the normalized channel estimate, 
$\sigma_{\widetilde{\pmb{H}}}^2=\frac{1}{N_t K}\trace\boldsymbol{R}_{\widetilde{\pmb{H}}} $ is the variance on the channel estimation error, where the error covariance matrix is \textcolor{black}{$\boldsymbol{R}_{\widetilde{\pmb{H}}}=\mathbb{E}\left[\operatorname{vec}(\widetilde{\pmb{H}})^* \operatorname{vec}(\widetilde{\pmb{H}})^T\right]$}
and $\sigma_{\widehat{\pmb{H}}}^2 $ is the variance on the channel estimate.
Adopting an \ac{MMSE} channel estimator, then under the  orthogonality principle for \ac{MMSE} estimates, we can use $\sigma_{\widehat{\pmb{H}}}^2=1-\sigma_{\widetilde{\pmb{H}}}^2$, which in this case, allows us to re-write the effective \ac{SNR} as 
$
	\frac{\sigma_{\widehat{\pmb{H}}}^2}{1+ \sigma_{\widetilde{\pmb{H}}}^2}=
	\frac{2}{1+ \sigma_{\widetilde{\pmb{H}}}^2}-1.
$
Hence, maximizing the worst case channel capacity can be achieved by minimizing $\sigma_{\widetilde{\pmb{H}}}^2$, i.e. minimizing the trace of error covariance matrix of the channel estimate, i.e. $  \operatorname{tr}\boldsymbol{R}_{\widetilde{\pmb{H}}}$.

%% file: Actions/motivation-MI-for-comm-part2.tex
Interestingly, for our adopted communication \ac{MI}, we have the following

\textbf{Theorem 2}: \textit{For a given channel estimation procedure giving $\widehat{\boldsymbol{h}}_k$ by observing $\pmb{y}_k$, the quantity $\mathcal{M}_k^{\tt{comm}}(\pmb{\Phi})$ can be lower-bounded as follows}
\begin{equation}
\label{eq:MI-for-comm-3}
	\mathcal{M}_k^{\tt{comm}}(\pmb{\Phi})
	 \geq h(\pmb{h}_k\vert \pmb{\Phi})-N_t \log(\pi e) -  N_t\log \frac{\operatorname{tr}(\boldsymbol{R}_{\widetilde{\boldsymbol{h}}_k})}{N_t} ,
\end{equation}
\textit{where $\widetilde{\boldsymbol{h}}_k=\boldsymbol{h}_k-\widehat{\boldsymbol{h}}_k$ and \textcolor{black}{$\boldsymbol{R}_{\widetilde{\boldsymbol{h}}_k}=\mathbb{E}\left[\widetilde{\boldsymbol{h}}_k \widetilde{\boldsymbol{h}}_k^H\right]$}.}

\textbf{Proof}: See \textbf{Appendix \ref{app:lower-bound-Mcomm}}.

\textcolor{black}{\input{Actions/fano-converse.tex}}
In fact, the variance on the individual channel estimation errors, i.e. $\sigma_{\widetilde{\pmb{h}}_k}^2=\frac{1}{N_t }\trace\boldsymbol{R}_{\widetilde{\boldsymbol{h}}_k}$ is easily verified to relate $\sigma_{\widetilde{\pmb{H}}}^2$ as $\sigma_{\widetilde{\pmb{H}}}^2 = \sum\nolimits_{k=1}^K \sigma_{\widetilde{\pmb{h}}_k}^2/K$.
From here, a decrease on the variance of channel estimation of the $k^{th}$ user, i.e. $\sigma_{\widetilde{\pmb{h}}_k}^2$, can contribute to an increase of the lower bound of $\mathcal{M}_k^{\tt{comm}}(\pmb{\Phi})$ according to \eqref{eq:MI-for-comm-3}.

%% file: Actions/fano-converse.tex
It is interesting to see that this lower bound can also be achieved by the estimation-theoretic counterpart of Fano's inequality. Indeed, applying \cite{cover1999elements} (c.f. Corollary of Theorem 8.6.6), one can obtain $\frac{1}{N_t}\mathbb{E}\left[\|\pmb{h}_k-\widehat{\pmb{h}}_k\|^2\right] \geq \frac{1}{\pi e} e^{\frac{1}{N_t}\left(h(\pmb{h}_k \mid \pmb{\Phi})-\mathcal{M}_k^{\tt{comm}}(\pmb{\Phi})\right)}$. After re-arrangement of terms, we arrive at \eqref{eq:MI-for-comm-3}.

%% file: Actions/motivation-MI-for-comm-part3.tex
The same decrease $\sigma_{\widetilde{\pmb{h}}_k}^2$ also decreases $\sigma_{\widetilde{\pmb{H}}}^2 $, which translates to a direct increase in $C_{\text {worst }}$.
This means that, through the intermediate variable $\sigma_{\widetilde{\pmb{h}}_k}^2$, any decrease of $\sigma_{\widetilde{\pmb{h}}_k}^2$ can simultaneously increase $\mathcal{M}_k^{\tt{comm}}(\pmb{\Phi})$, as well as $C_{\text {worst }}$.
To illustrate this influence, we have plotted in Fig. \ref{fig:MIcomm_vs_Cworst} how the behaviour of $\mathcal{M}^{\tt{comm}}(\pmb{\Phi})$ varies with the worst case channel capacity $C_{\text {worst }}$ given in \eqref{eq:Cworse}. 
It is evident that an increase of worst case channel capacity leads to the direct increase of $\mathcal{M}^{\tt{comm}}(\pmb{\Phi})$, in the long-term.
It is worth noting that we have generated random pilot matrices and used them to perform \ac{MMSE} channel estimation in order to empirically compute \eqref{eq:Cworse}. We have used these same pilot matrices to compute the \ac{MI} metric.
The increasing trend is also witnessed on varying communication noise levels.

%% file: Actions/motivation-for-sensing.tex
A question that one may ask is \textit{"How does the sensing \ac{MI} adopted in this paper relate to any detection performance metrics ?"}
In this Section, we show that $\mathcal{M}^{\text {sense }}(\boldsymbol{\Phi})$ is asymptotically related to the probability of detection of the most powerful detection test, according to Neyman-Pearson criterion, under a fixed probability of false alarm.
In addition, we also answer a question in the context of parameter estimation, that is \textit{"Is there any connection between the considered sensing \ac{MI} metric and existing estimation performance metrics, such as \ac{MSE} or \ac{CRB} ?"}
\paragraph{Detection tasks} For sensing, this paper focuses on detection rather than estimation. For this, we have the following theorem

\textbf{Theorem 3}: \textit{For the most powerful hypothesis test, the function $\mathcal{M}^{\tt{sense}}(\pmb{\Phi})$ relates to the probability of detection as}
\begin{equation}
\label{eq:MI-for-sense-mot}
\lim _{L \rightarrow \infty}\left(-\frac{1}{L} \log \left(1-P_{\mathrm{D}}\right)\right)
	=\mathcal{M}^{\tt{sense}}(\pmb{\Phi}) - g(\boldsymbol{\nu},\pmb{\Phi}),
\end{equation}
\textit{where $g(\boldsymbol{\nu},\pmb{\Phi}) = \frac{\nu_0 \boldsymbol{\mu}_0^H\left(\sum_{i=1}^Q \nu_i \boldsymbol{\mu}_i \boldsymbol{\mu}_i^H+\sigma_r^2 \boldsymbol{I}\right)^{-1} \boldsymbol{\mu}_0}{1+\nu_0 \boldsymbol{\mu}_0^H\left(\sum_{i=1}^Q \nu_i \boldsymbol{\mu}_i \boldsymbol{\mu}_i^H+\sigma_r^2 \boldsymbol{I}\right)^{-1} \boldsymbol{\mu}_0} \geq 0 $ and $\boldsymbol{\nu}$ is a vector of $\nu_0 \ldots \nu_Q$.}

\textbf{Proof}: See \textbf{Appendix \ref{app:Stein-on-Msense}}.

The above theorem tells us that $\mathcal{M}^{\tt{sense}}(\pmb{\Phi})$ plays a crucial role, asymptotically with the number of samples and in the large radar \ac{SCNR} regime.
To be more specific, in the large radar \ac{SCNR} regime, i.e. when $\nu_0$ grows large with respect to $\nu_1 \ldots \nu_Q$ and $\sigma_r^2$, notice that $g(\boldsymbol{\nu},\pmb{\Phi})$ converges to $1$, i.e. it becomes independent of $\pmb{\Phi}$.
In other words, a linear increase in $P_{\mathrm{D}}$ contributes to a logarithmic increase in $\mathcal{M}^{\tt{sense}}(\pmb{\Phi})$ within the prescribed asymptotic regime.
This is to say that optimizing $\mathcal{M}^{\tt{sense}}(\pmb{\Phi})$ can influence the detection performance.
Within this context, we also note that a result in \cite{87006} highlights that higher \ac{MI} between a parameter and its measurement leads to improved expected performance in the optimal Bayes risk decision procedure.

\paragraph{Estimation tasks} Estimation as a sensing task can be also discussed in this Section. 
Let's say now that the \textcolor{black}{\ac{ISAC}} \ac{BS} aims at estimating $\pmb{\xi} = \begin{bmatrix}
	\pmb{\Theta} ,&
	\pmb{\nu}
\end{bmatrix}^T$, rather than detecting the presence of a target. 
Following a similar procedure as the proof of \textbf{Theorem 2}, one can lower bound $\mathcal{M}^{\tt {sense }}(\pmb{\Phi})$ as 
\begin{equation}
\label{eq:M_sense_low_bound}
\mathcal{M}^{\tt {sense }}(\pmb{\Phi}) \geq 
	h\left(\pmb{\xi} \mid \boldsymbol{\Phi}\right)-\log \left(\operatorname{det}\left(\pi e \boldsymbol{R}_{\tilde{\pmb{\xi}}}\right)\right)
\end{equation}
where \textcolor{black}{$\tilde{\pmb{\xi}} = \widehat{\pmb{\xi}}-{\pmb{\xi}}$ and $\boldsymbol{R}_{\tilde{\pmb{\xi}}}=$ $\mathbb{E}\left[\tilde{\pmb{\xi}}\tilde{\pmb{\xi}}^H\right]$}, which tells us that decreasing the error, in the sense of \ac{MSE}, translates to an increase in the lower bound of the sensing \ac{MI}.
We are also able to relate the sensing \ac{MI} with the \ac{CRB}. The \ac{CRB} inequality tells us that, under the condition that $\widehat{\pmb{\xi}}$ is an unbiased estimator of ${\pmb{\xi}}$, we have that
\begin{equation}
	\boldsymbol{R}_{\tilde{\pmb{\xi}}} \succeq \pmb{\mathcal{F}}^{-1}=\left(E\left\{\left[\nabla \ln \ell\left(\boldsymbol{Y}_r ; \pmb{\xi}\right)\right]\left[\nabla \ln \ell\left(\boldsymbol{Y}_r ; \pmb{\xi}\right)\right]^H\right\}\right)^{-1},
\end{equation}
where $\pmb{\mathcal{F}}$ is the \ac{FIM} and, subsequently, $\pmb{\mathcal{F}}^{-1}$ is the \ac{CRB} matrix. Also, $\ell()$ is the likelihood function. Since $\pmb{\mathcal{F}}^{-1}$ is an error covariance matrix itself, it can replace $\boldsymbol{R}_{\tilde{\pmb{\xi}}}$ in \eqref{eq:M_sense_low_bound} and so
\begin{equation}
	\mathcal{M}^{\tt {sense }}(\boldsymbol{\Phi}) \geq h\left(\pmb{\xi} \mid \boldsymbol{\Phi}\right)-N \log \pi e+\log (|\boldsymbol{\mathcal{F}}|).
\end{equation}
It is worth noting that many previous works have adopted the maximization of the \ac{FIM} in the determinant sense, such as \cite{4359542},\cite{10483086}.
Typically, a higher sensing \ac{MI} allows us to extract more information about the target from the collected measurements, \textit{as though the target were directly communicating its sensing parameters with the \textcolor{black}{\ac{ISAC}} \ac{BS}.}

%% file: Actions/motivation-unify.tex
It is tempting to consider a common metric for both sensing and communication, rather than the common way of considering metrics of different nature (e.g. \ac{SINR} for communications, and \ac{CRB} or detection probability for sensing).
% same unit
Under the proposed framework, one can treat the overall weighted objective between sensing and communications, given in \eqref{eq:scalarization1}, as a metric under a common unit of measurement, as both the \ac{MI}-based metric for sensing and that for communications are both measured in $\operatorname{bits}$. The proposed framework of \ac{MOOP} with a Pareto order relation defined on the $\mathbb{R}^{K+1}$ Euclidean space offers a formal way of combining the \ac{MI}-based metrics for sensing and communications. 
Thanks to such a unification, we can extract the overall bits of information for sensing, in terms of detection, and communications, in terms of channel estimation, where the latter has been shown in Section \ref{subsec:MI-for-comm} to impact channel capacity.
Also, the proposed projected gradient method described in \eqref{eq:GD-step}, \eqref{eq:project-step}, \eqref{eq:GD-project} is computationally efficient, thanks to the combined \ac{MI} metrics.

%% file: sections/simulation-results.tex
In this section, we carry out simulations to demonstrate the \ac{MI}-based \ac{ISAC} tradeoffs, as well as the performance advantages of the proposed pilot design method. The array configuration at the \textcolor{black}{\ac{ISAC}} \ac{BS} follow a \ac{ULA} fashion for both transmit and receive arrays as shown given in \eqref{eq:steering-1} and \eqref{eq:steering-2}. The central frequency is set to $f_c = 5 \GHz$.
We take $N_k = 180$ over all $K$ communication users. 
Following \cite{saunders2007antennas}, the \ac{PDF} of each channel component follows a \ac{GMM} given in \eqref{eq:GMM}, where each activation probability is modeled through a Laplacian distribution, i.e. $\alpha_{k,n} = \frac{1}{\sqrt{2}\bar{\sigma}_k}\exp(-\frac{\sqrt{2}\vert \theta_n -  \bar{\theta}_k \vert}{\bar{\sigma}_k})$, where $\bar{\sigma}_k$ is the azimuth spread characterizing the channel between the \textcolor{black}{\ac{ISAC}} \ac{BS} and the $k^{th}$ communication user. Also, $\bar{\theta}_k$ represents the mean \ac{AoA} of the spread of the channel towards the $k^{th}$ user. 
Each Gaussian component of the \ac{GMM} is assumed to have a covariance matrix $\pmb{R}_{k,n} = \int_{\mathcal{R}_n} \pmb{a}_{N_t}(\theta) \pmb{a}_{N_t}^H(\theta) \ d\theta $, where $\mathcal{R}_n$ is the region of over which the $n^{th}$ \ac{GMM} component is assumed to be observed. These regions are assumed to be disjoint and $\bigcup_n \mathcal{R}_n$ covers $[-90^\circ,+90^\circ]$. Therefore, to generate a random realization of channel $\pmb{h}_{k}$, we first generate $N_k = 180$ Gaussian random realizations, where the $n^{th}$ component has mean $\pmb{\mu}_{k,n}$ and co-variance matrix $\pmb{R}_{k,n}$, then $\pmb{h}_{k}$ is picked based on the activation probabilities $\alpha_{k,n}$.
For the single-user case, i.e. $K  = 1$, we set the mean \ac{AoA} component to $\bar{\theta}_1 = 70^\circ$. 
For the multi-user case, where we have set $K = 4$, we assume that $\bar{\theta}_2 = 23^\circ$, $\bar{\theta}_3 = -23^\circ$,  and $\bar{\theta}_4 = - 70^\circ$. All communication users are treated equally, therefore $w_k = \frac{1}{K}$.
Unless otherwise stated, the projected gradient descent algorithm runs with a step size of $\gamma = 0.1$.
\textcolor{black}{\input{Actions/R2C5}}

%%%%%%%%%%%%%% MI tradeoff (N,L) no clutter %%%%%%%%%%%%%%
\begin{figure}[t]
	\centering
	\includegraphics[width=1\linewidth]{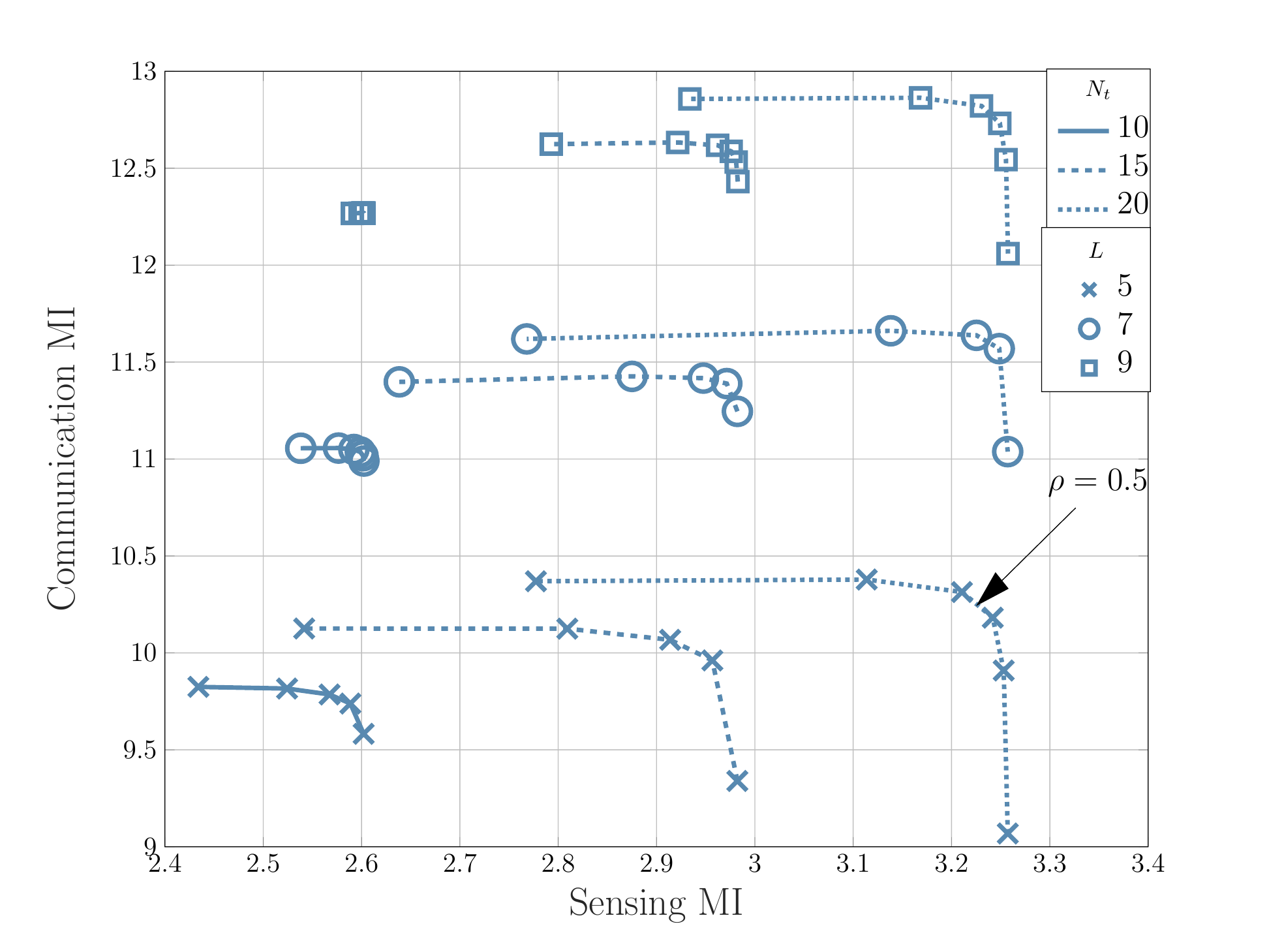}
	\caption{Achievable sensing and communication \ac{MI} values for the single-user case by projected gradient descent optimization for different values of $N_t$ and $L$. The number of receive antennas is set to $N_r = 5$. The target is located at $\theta_0 = 60^\circ$. The radar noise is $\sigma_r = 2$ and the communication noise is $\sigma_1 = 0.1$. No clutter is considered here.}
	\label{fig:MI}
\end{figure}

\begin{figure}[t]
	\centering
	\includegraphics[width=1\linewidth]{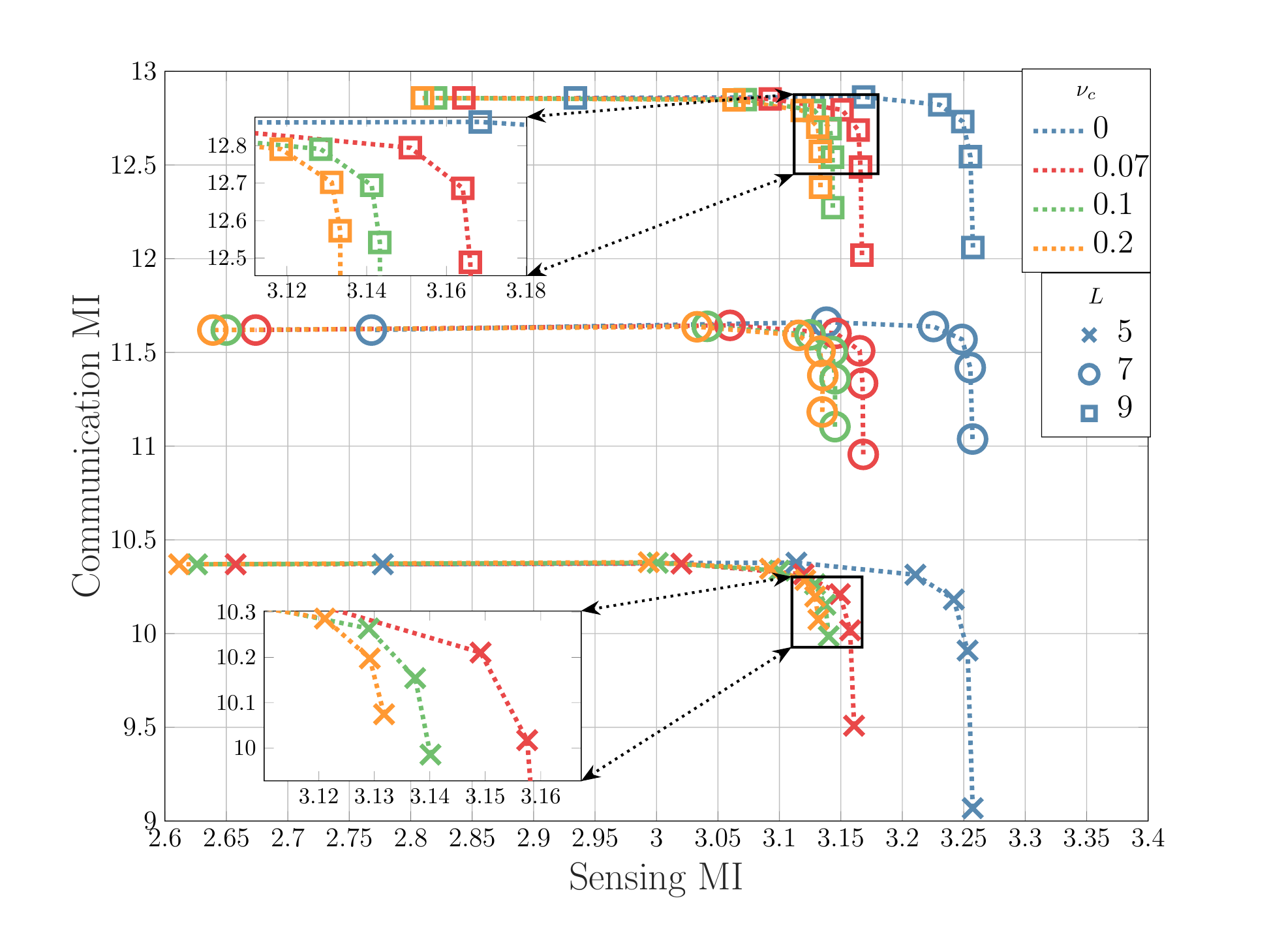}
	\caption{Clutter impact on the \ac{MI} \ac{ISAC} performance for the single-user case and different clutter power $\nu_c$ and values of $L$. We set $\theta_1 = 0^\circ$, $N_r = 5$, $\theta_0 = 60^\circ$, $\sigma_r = 2$ and $\sigma_1 = 0.1$.}
	\label{fig:MI_clutter}
\end{figure}
\paragraph{Impact of $N_t$ and $L$ on the \ac{MI} \ac{ISAC} trade-off}
In Fig. \ref{fig:MI}, we study the impact of number of transmit antennas, $N_t$ and number of pilot symbols (pilot length) $L$ on the sensing and communication mutual information. For fixed $L$ and with increasing number of transmit antennas, we can see that both sensing and communication mutual information improve. For example, fixing $L = 5$, we see that maximum communication \ac{MI} improves from $9.82$ bits to $10.12$ bits when increasing $N_t$ from $10$ to $15$ antennas, while the maximum sensing \ac{MI} rises from $2.6$ to $2.98$ bits. The same observation is noticed when going from $N_t = 15$ to $N_t = 20$ for $L = 5$. We set $\sigma_r = 2$. We assume that the single-user communication noise is $\sigma_1 = 0.1$. Furthermore, the azimuth spread is set to $\bar{\sigma}_1 = 6^\circ$.
Moreover, the target is located at $\theta_0 = 60^\circ$, and the number of receive antennas is set to $N_r = 5$. Another point worth highlighting is the point corresponding to $\rho = 0.5$, where we have equal priority on communication and sensing performance. We also observe that increasing $N_t$ also contributes to a simultaneous increase in both communication and sensing \ac{MI}. For $\rho = 0.5$, we observe that for $L = 5$, the sensing \ac{MI} increases from $2.56$ bits to $3.21$ bits, while the communication \ac{MI} increases from $9.78$ bits to $10.31$ bits, when doubling the number of transmit antennas from $10$ to $20$.
Another factor contributing to the joint \ac{ISAC} \ac{MI} gain is $L$. For instance, focusing on the point corresponding to $\rho = 0.5$ and fixing $N_t = 15$, we observe that the sensing \ac{MI} increases from $2.91$ bits to $2.99$ bits, whereas the communication \ac{MI} increases from $10.06$ bits to $12.58$ bits, when $L$ increases from $5$ to $9$. 
An interesting phenomenon worth highlighting is the achievable range of \ac{ISAC} tradeoffs, as a function of $N_t$ and $L$. For example, when $L = 9$, we notice that increasing $N_t$ not only improves the \ac{ISAC} \ac{MI}, but also allows the designer to achieve a wider set of sensing and communication \ac{MI} values for different \ac{ISAC} pilot matrices. Indeed, notice that for $N_t = 10$ and $L = 9$, the trade-off achieves sensing \ac{MI} values that are within $[2.59,2.61]$, whereas communication \ac{MI} values are within $[12.265,12.27]$. Both range of values can be increased by increasing $N_t$, thus allowing for more \ac{ISAC} trade-offs. 

%%%%%%%%%%%%%% MI tradeoff (L) with clutter %%%%%%%%%%%%%% 
\paragraph{Impact of clutter on the \ac{MI} \ac{ISAC} trade-off}

In Fig. \ref{fig:MI_clutter}, we highlight how clutter can influence the \ac{ISAC} performances from a mutual information perspective.
 The same simulation parameters as in Fig. \ref{fig:MI}, with the exception of the clutter setting. In particular, to study the impact of clutter, we set $Q = 1$ and denote $\nu_c = \nu_1$. We vary the values of $\nu_c$ to analyze the clutter impact. The clutter is assumed to be located at $\theta_1 = 0^\circ$.
 We can see that for fixed $N_t$ and $L$, an increase in $\nu_c$ causes the sensing \ac{MI} to decrease without majorly impacting the communication \ac{MI}. Despite the presence of clutter, we see that the sensing \ac{MI} is still relatively high enough, hence detection can still be reliably performed. Fixing $L$ to $5$, we see that a clutter component with $\nu_c = 0.07$ deteriorates the maximal sensing \ac{MI} from $3.25$ to $3.16$ bits compared to the no-clutter case. On the other hand, the maximal communication \ac{MI} is at $10.37$ bits regardless of the clutter power. When equal priority is set on sensing and communication, increasing $L$ can aid in improving the sensing performance. For example, the sensing \ac{MI} increases from $3.14$ to $3.16$ bits when increasing $L$ from $5$ to $9$. Similar observations can be reported for increasing $\nu_c$. 

%%%%%%%%%%%%%% MI tradeoff when target approaches user %%%%%%%%%%%%%% 
\begin{figure}[t]
	\centering
	\includegraphics[width=1\linewidth]{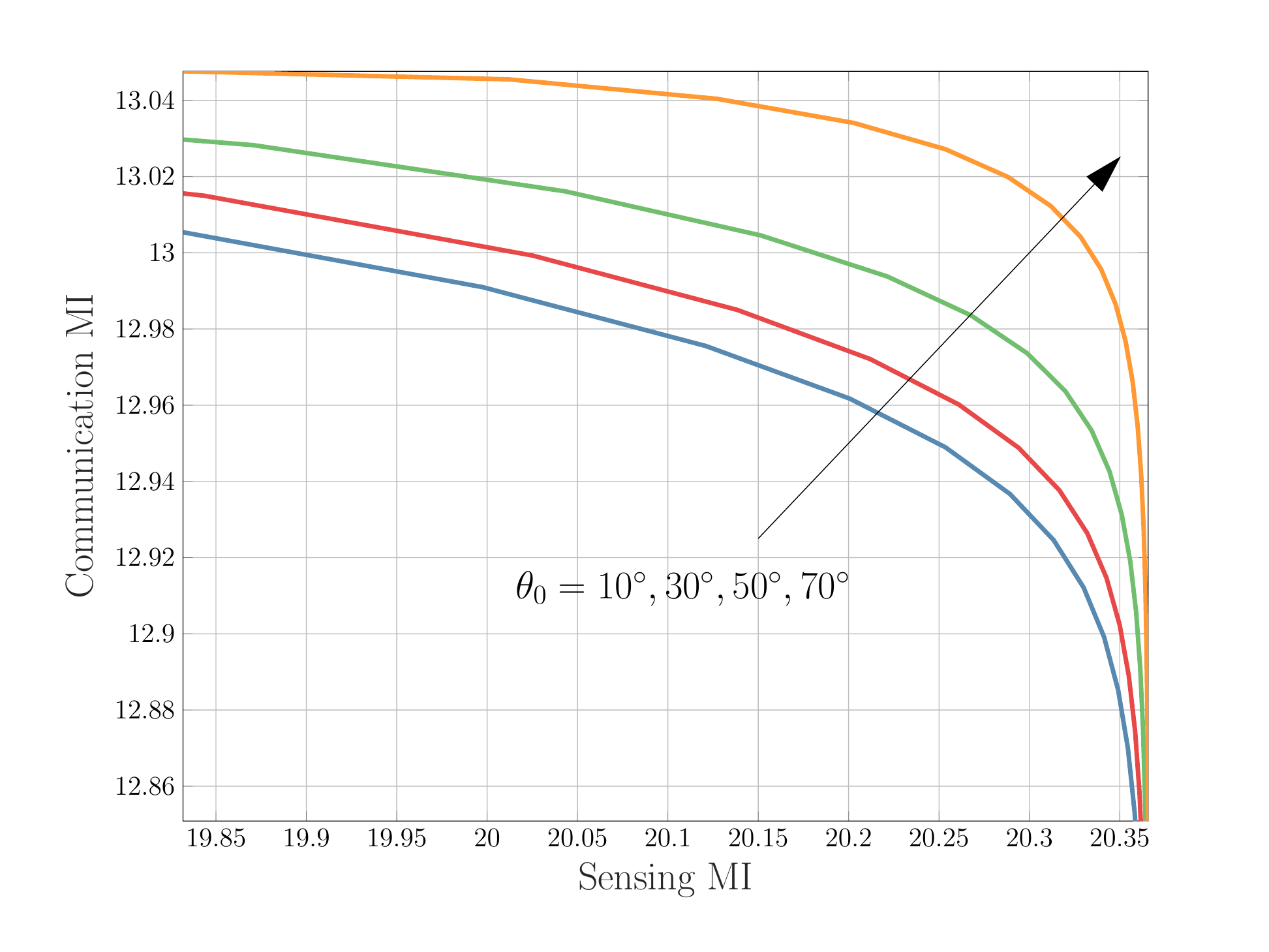}
	\caption{The impact of the target \ac{AoA} on the \ac{ISAC} trade-off for the single-user case. We plot $4$ \ac{MI}-frontiers where the blue frontier corresponds to $\theta_0 = 10^\circ$ and the orange corresponds to $\theta_0 = 70^\circ$. We set $N_r = 70$, $L = 10$. We have $\sigma_1 = 0.1$ and $\sigma_r = 0.002$. The user is located around $\bar{\theta}_1 = 70^\circ$.}
	\label{fig:MI_target_to_user}
\end{figure}
\paragraph{Target location can improve \ac{MI} \ac{ISAC} gains}
In Fig. \ref{fig:MI_target_to_user}, we aim at studying the target position and its impact on the sensing and communication \ac{MI} tradeoffs.
The single-user case is assumed with $L = 10$ and number of receive antennas $N_r = 70$ at the \textcolor{black}{\ac{ISAC}} \ac{BS}. We set $\sigma_1 = 0.1$ and $\sigma_r = 0.002$.
The frontiers are generated by sweeping over multiple values of $\rho$. The different frontiers are generated by changing the target's location.
The user is located around $\bar{\theta}_1 = 70^\circ$.
In particular, we vary the target \ac{AoA} by bringing it closer towards the mean of the communication user \ac{AoA}. Indeed, the closer the target is brought towards the user, the boundary approach a utopia point, i.e. achieving maximal sensing and communication \ac{MI} performance with the same orthogonal pilot symbols. 
This can be explained by some sensing and communication \textit{information overlap} between the communication channel $\pmb{h}_k$ and part of the sensing channel characterized by the \ac{AoA}s $\pmb{\Theta}$ and the path gains $\pmb{\nu}$.
Note that a similar phenomenon is reported in \cite{10147248}, where a sensing and communication \textit{subspace overlap} can also improve communication-rate and sensing performance. The study, herein, serves as a complementary one as pilots can provide better channel estimates when the target approaches the communication user. The best performance is observed when the target and communication user are at the same location, which is an interesting use case when the objective is to sense a communication user, hence the integration gain is fully exploited. 

%%%%%%%%%%%%%% ROC 1 %%%%%%%%%%%%%% 
\begin{figure}[t]
	\centering
	\includegraphics[width=1\linewidth]{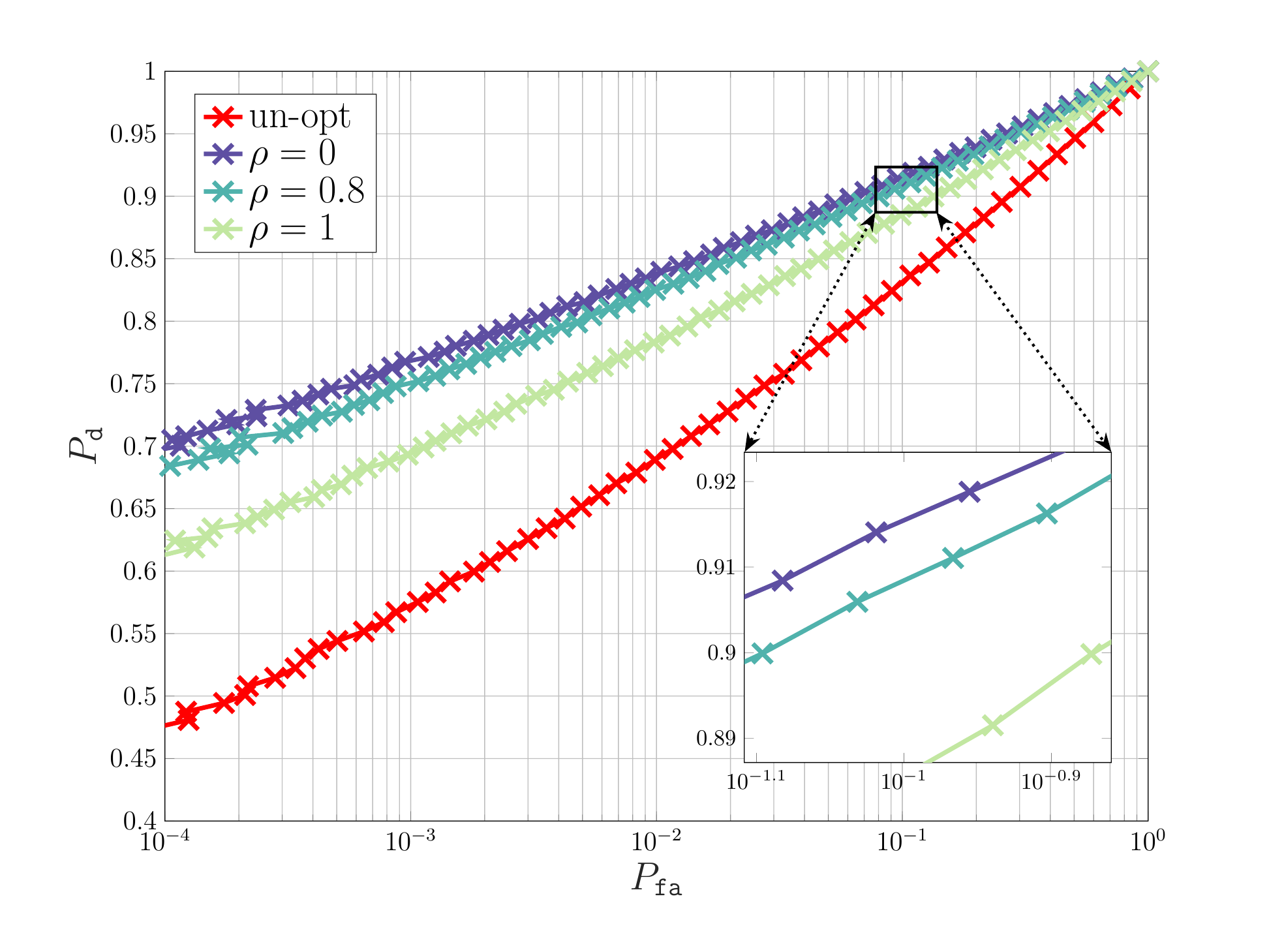}
	\caption{Receiver operating characteristic curves for different values of $\rho$ and for an non-optimized and orthogonal pilot, i.e. random but orthogonal, for $\sigma_r = 2$. The single-user case is considered with $N_t = 20$ and $N_r = 5$ antennas. The pilot length is $L = 9$. The target is located at $\theta_0 = 60^\circ$ and $\sigma_1 = 0.1$.}
	\label{fig:ROC_1}
\end{figure}
\paragraph{ROC performance of pilot matrices}
\label{paragraph:ROC-performance}
\textcolor{black}{\input{Actions/mentioned-ROC-monte-carlo}}In Fig. \ref{fig:ROC_1}, we take a step further and util	ize the generated pilot symbols to evaluate the performance of the \ac{ROC} via the optimal detector reported in  \cite{kay1993fundamentals}. In particular, the figure shows the \ac{ROC} curves for different values of $\rho$ and compares the \ac{ROC} performance to an non-optimized and orthogonal pilot matrix. To this end, we set the pilot length is $L = 9$. Moreover, the number of transmit and receive antennas at the \textcolor{black}{\ac{ISAC}} \ac{BS} are set to $N_t = 20$ and $N_r = 5$ antennas, respectively. 
It is clear that optimizing the pilot matrix via the proposed projected gradient method can improve the detection capabilities performed at the \textcolor{black}{\ac{ISAC} \ac{BS}} through the same pilot matrix used for communications. For this simulation,  the target is located at $\theta_0 = 60^\circ$ and the communication noise level is $\sigma_1 = 0.1$.
As an example, if the designer sets a probability of false alarm at $P_{\tt{fa}} = 10^{-4}$ and generates an orthogonal pilot matrix by the proposed method for $\rho = 1$ (communication-optimal), the achieved probability of detection is $P_{\tt{d}} = 0.61$, which is greater than that of an non-optimized orthogonal pilot, which achieves $P_{\tt{d}} = 0.475$. Note that, even though the priority is set to the communication task (i.e. channel estimation), the optimized pilot generated by the proposed method can still outperform an non-optimized one. If the designer seeks a better detection performance, then this can be achieved by lowering $\rho$, hence giving more priority towards sensing. Indeed, with $\rho = 0.8$ and for the same $P_{\tt{fa}} = 10^{-4}$, the $P_{\tt{d}}$ increases from $0.61$ to $0.68$ and can reach $P_{\tt{d}} = 0.7$ for $\rho = 1$. Similar observations are noticed for any false-alarm probability.  
%%%%%%%%%%%%%% ROC 2 %%%%%%%%%%%%%% 
\begin{figure}[t]
	\centering
	\includegraphics[width=0.9\linewidth]{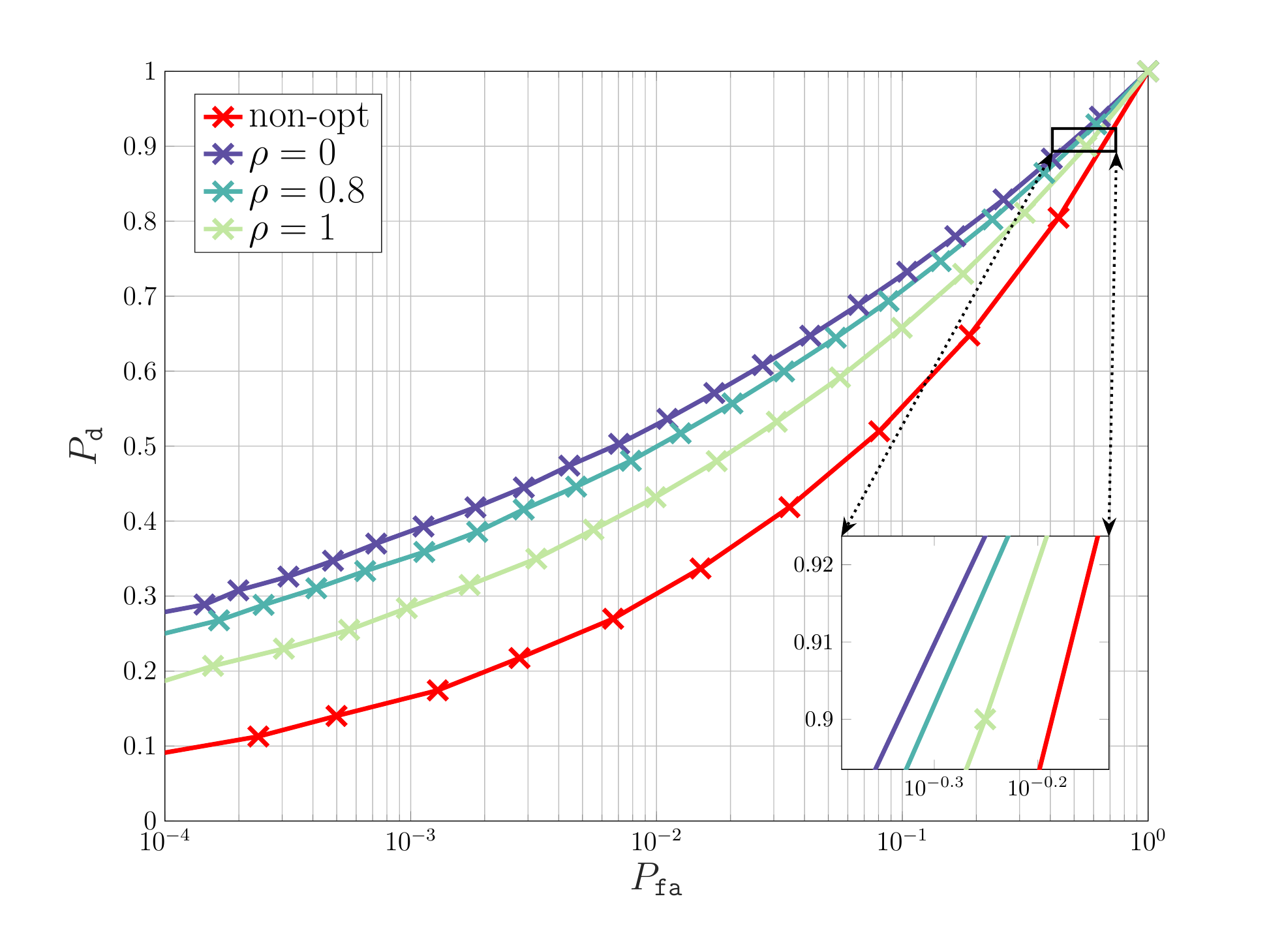}
	\caption{Receiver operating characteristic curves for different values of $\rho$ and for an non-optimized and orthogonal pilot for $\sigma_r = 4$. All other simulation parameters are the same as those in Fig. \ref{fig:ROC_1}.}
	\label{fig:ROC_2}
\end{figure}
\begin{figure}[t]
	\centering
	\includegraphics[width=0.8\linewidth]{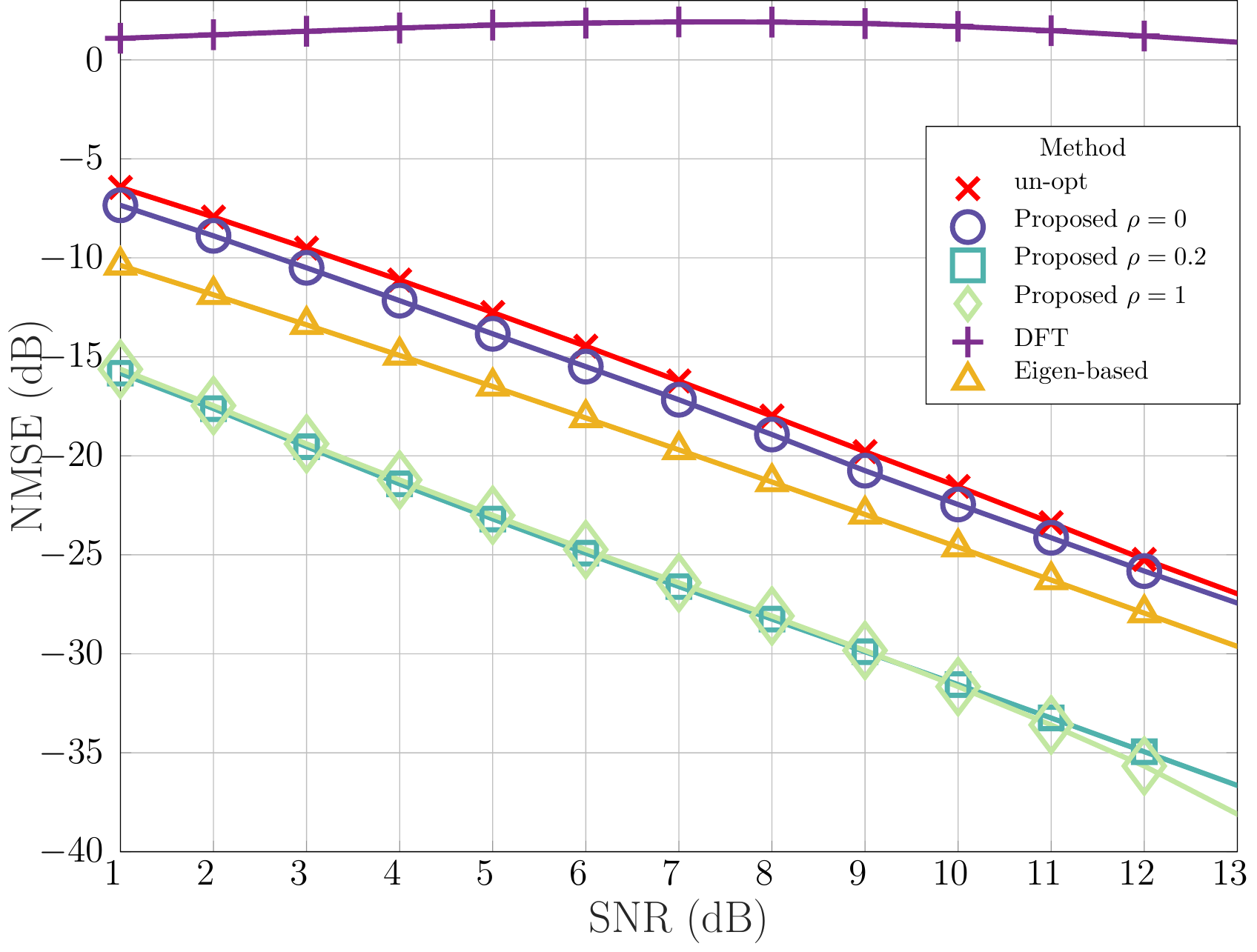}
	\caption{\textcolor{black}{NMSE performance of the generated pilots compared to an unoptimized orthogonal pilot matrix for different number of communication users and different values of $\rho$.}}
	\label{fig:NMSE}
\end{figure}

In Fig. \ref{fig:ROC_2}, we increase the radar noise to $\sigma_r = 4$. All other simulation parameters are exactly the same as those in Fig. \ref{fig:ROC_1}. Indeed, due to a more noisy radar sub-system, the detector has to increase the probability of false alarm to achieve the same detection probability performance as those specified in Fig. \ref{fig:ROC_1} (i.e. $\sigma_r = 2$) for $P_{\tt{d}} = 0.475$. Given that, increasing the $P_{\tt{fa}}$ level to $P_{\tt{fa}}=5 \times 10^{-2}$, the non-optimized pilot achieves $P_{\tt{d}} = 0.475$, whereas the proposed method for $\rho = 1$ generates an orthogonal pilot with $P_{\tt{d}} = 0.59$ capabilities. If a larger detection probability is desired, one would have to decrease $\rho$ thus achieving $P_{\tt{d}} = 0.64$ for $\rho = 0.8$ and $P_{\tt{d}} = 0.67$ for $\rho = 1$. We have identical observations for different false-alarm levels.

%%%%%%%%%%%%%% NMSE %%%%%%%%%%%%%% 

\paragraph{NMSE performance} 
In Fig. \ref{fig:NMSE}, we assess the \ac{NMSE} performance of the generated orthogonal pilot matrices in order to study its channel estimation capability with different number of communication users, i.e. $K$, and different values of $\rho$. To this end, the \ac{NMSE} is evaluated as 
\begin{equation}
	\NMSE 
	=
	\frac{1}{KR}
	\sum\nolimits_{k = 1}^K
	\sum\nolimits_{r = 1}^R
	\frac{\Vert \pmb{h}_{k,r} - \widehat{\pmb{h}}^{\tt{MMSE}}_{k,r} \Vert^2}{\Vert \pmb{h}_{k,r} \Vert^2},
\end{equation}
where $\pmb{h}_{k,r} $ is the communication channel between the \textcolor{black}{\ac{ISAC}} \ac{BS} and the $k^{th}$ user generated at the $r^{th}$ Monte-Carlo trial, and $\widehat{\pmb{h}}^{\tt{MMSE}}_{k,r} $ is its \ac{MMSE} estimate, which is a widely adopted figure of merit in signal processing \cite{guo2013interplay}. As a reminder, the \ac{MMSE} estimate minimizes $\textcolor{black}{\mathbb{E}[ \Vert \pmb{h}_{k,r} - \widehat{\pmb{h}}_{k,r} \Vert^2 ]}$ with respect to $\widehat{\pmb{h}}_{k,r} $ and is obtained via the following marginalization as follows \cite{fesl2022channel}
\begin{equation}
\label{eq:MMSE-1}
\begin{split}
	\widehat{\pmb{h}}^{\tt{MMSE}}_{k,r} 
	&\triangleq
	\textcolor{black}{\mathbb{E}[ {\pmb{h}}_{k,r} \vert \pmb{y}_{k,r} ]} \\
	&=
	\sum\limits_{n=1}^{N_k}
	p_{k,n,r}
	(\pmb{\mu}_{k,n} + \pmb{R}_{k,n} \pmb{\Phi}^H \pmb{\Sigma}^{-1}_{k,n}(\pmb{\Phi}) (\pmb{y}_{k,r}-\pmb{\Phi}\pmb{\mu}_{k,n})),
\end{split}
\end{equation}
where $\pmb{y}_{k,r} \in \mathbb{C}^{L \times 1}$ is the received signal at the $k^{th}$ user on the $r^{th}$ trial following \eqref{eq:comm-model}. Moreover, $\pmb{\Sigma}_{k,n}(\pmb{\Phi}) $ is given in equation \eqref{eq:Sigma_Phi}. The probabilities at the $r^{th}$ trial $p_{k,n,r}$ (also referred to as \textit{responsibilities}) are computed via
\textcolor{black}{
\begin{equation}
\label{eq:MMSE-2}
	p_{k,n,r} = \frac{\alpha_{k,n}f_{\mathcal{CN}}(\pmb{y}_{k,r} \vert\pmb{\Phi}\pmb{\mu}_{k,n}, \pmb{\Sigma}_{k,n}(\pmb{\Phi}))}{\sum_{n'=1}^{N_k} \alpha_{k,n'} f_{\mathcal{CN}}( \pmb{y}_{k,r}\vert\pmb{\Phi}\pmb{\mu}_{k,n'}, \pmb{\Sigma}_{k,n'}(\pmb{\Phi}))}.
\end{equation}
}
\textcolor{black}{\input{Actions/Addedbenchmarks1}}

%%%%%%%%%%%%%% SER %%%%%%%%%%%%%% 
\begin{figure}[t]
	\centering 
	\includegraphics[width=0.9\linewidth]{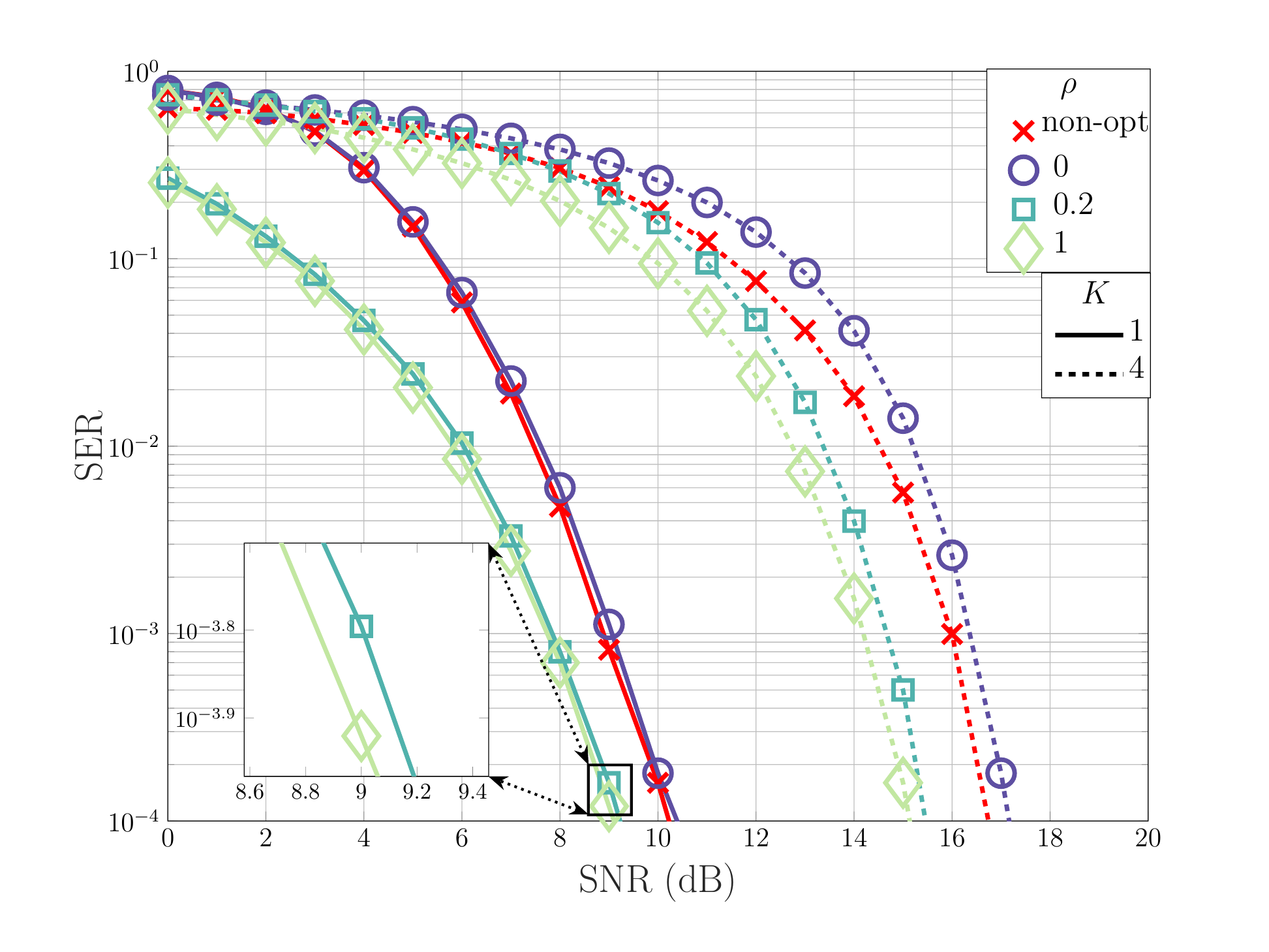}
	\caption{SER performance of the generated pilots compared to an unoptimized orthogonal pilot matrix for different values of $\rho$ and different number of communication users. A $64-$QAM with Gray encoding is used for data modulation. }
	\label{fig:SER}
\end{figure}
\paragraph{SER performance} In Fig. \ref{fig:SER}, we evaluate the \ac{SER} performance of the generated orthogonal pilot matrices and observe the gains in \ac{SER} as compared to the non-optimized pilot matrix as a function of \ac{SNR} and for different values of $\rho $ and number of communication users, i.e. $K$.
The same simulation parameters are used as those in Fig. \ref{fig:NMSE}. 
For communications, a $64-$\ac{QAM} was used as constellation for digital modulation with gray encoding.
The \ac{MMSE} channel estimator in \eqref{eq:MMSE-1} and \eqref{eq:MMSE-2} is first utilized to estimate the channel, then a \ac{ZF} equalizer is used to demodulate the symbol using a hard-decision decoder. 
For the single-user case, and setting the \ac{SER} level to $10^{-4}$, we see that the \ac{SER} performance of the non-optimized pilot matrix is $0.15\dB$ better than that of the pilot matrix generated by the proposed method for $\rho = 0$ (sensing-optimal). We can improve the \ac{SER} by $1\dB$ by increasing $\rho$ to $0.2$. An additional gain of about $0.1\dB$ can be achieved by increasing $\rho$ to $1$.
As for the multi-user case for $K = 4$, better gains can be reported, even though the non-optimized pilot gains $0.5\dB$ as compared to pilot matrix generated by the proposed method for $\rho = 0$. For example, setting the \ac{SER} level to $10^{-4}$ and $\rho = 0.2$, we can gain $1.3\dB$ of \ac{SNR} as compared to the non-optimized pilot and $1.6\dB$ if we fix $\rho = 1$. Therefore, we can report more \ac{SNR} gains with increasing number of communication users.
%%%%%%%%%%%%%% Iterations %%%%%%%%%%%%%% 
\begin{figure}[t]
	\centering
	\includegraphics[width=0.8\linewidth]{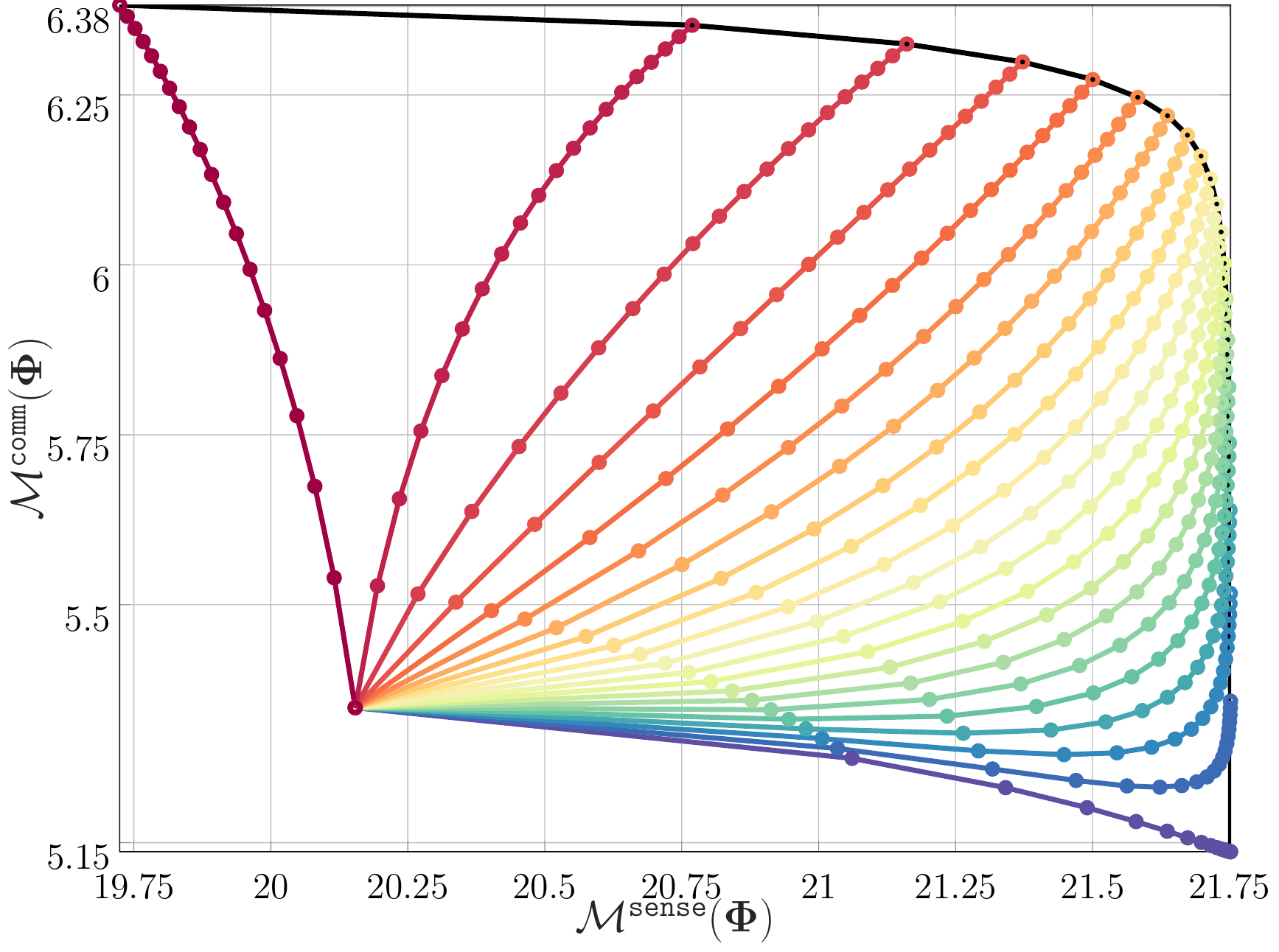}
	\caption{\textcolor{black}{The convergence behavior of the proposed method in terms of $\mathcal{M}^{\tt{comm}}(\pmb{\Phi})$ and $\mathcal{M}^{\tt{sense}}(\pmb{\Phi})$. Each colored path represents the path taken for a fixed $\rho$. Each point corresponds to a pilot matrix generated by the method. The initial point at $(20.15,5.34)$ is the initial point shared for all paths.} \textcolor{black}{The black line represents the frontier joining the converged points spanning $\rho = 0 \ldots 1$, i.e. an approximation of the Pareto frontier.}}
	\label{fig:Iter_1}
\end{figure}
\paragraph{Convergence \textcolor{black}{behaviour}}
\textcolor{black}{\input{Actions/updated_convergence_behaviour}}
%%%%%%%%%%%%%% Iterations %%%%%%%%%%%%%% 
\begin{figure}[t]
	\centering
	\includegraphics[width=0.8\linewidth]{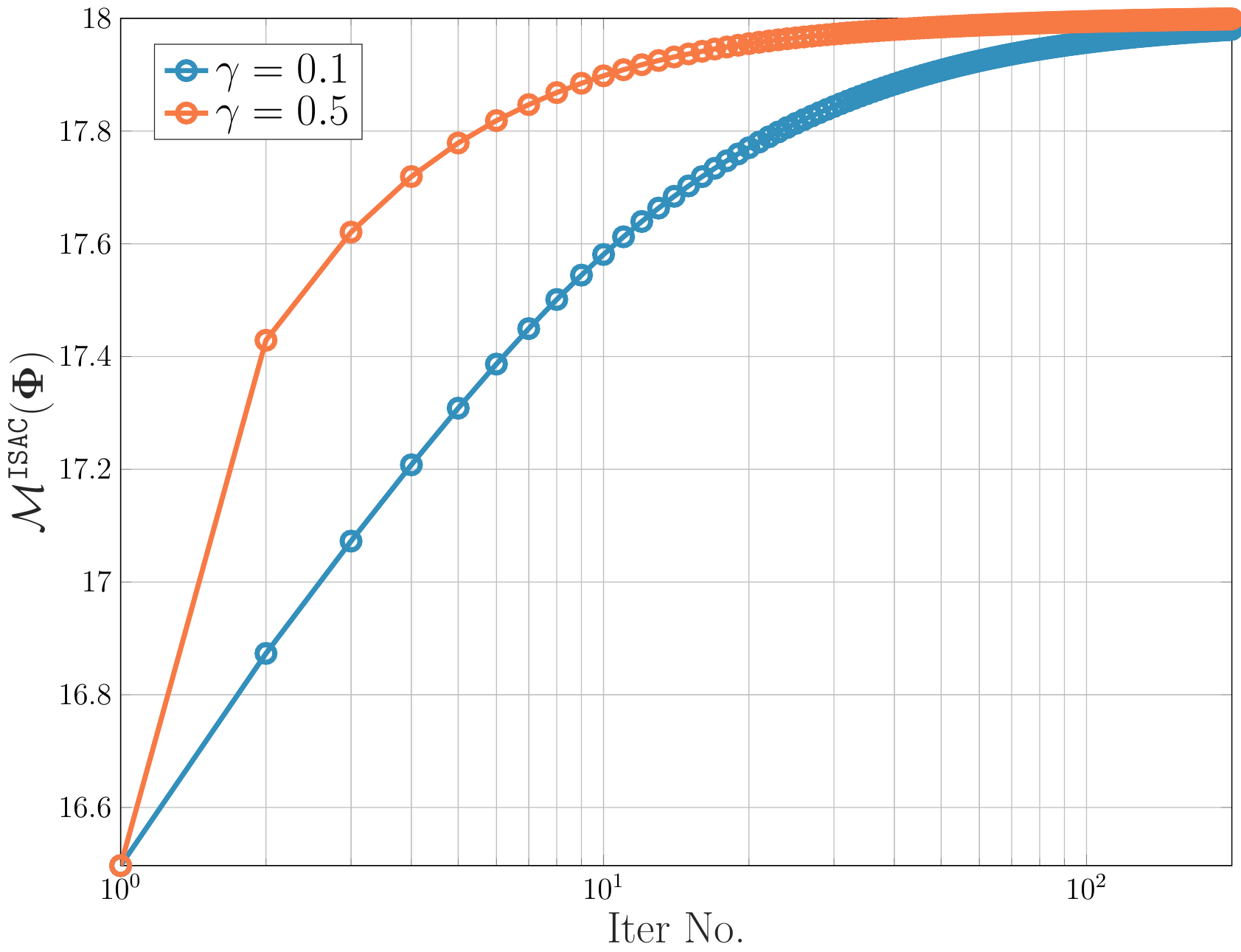}
	\caption{The impact of the step size on the convergence behavior of the projected gradient descent for \ac{ISAC} pilot design.}
	\label{fig:Iter_2}
\end{figure}

Next, in Fig. \ref{fig:Iter_2}, we plot the cost convergence versus the iteration number. We run the algorithm for $200$ iterations and for different values of step-size, $\gamma$. We see that for both values of $\gamma$, the method converges to a stable value of $\mathcal{M}^{\tt{ISAC}}(\pmb{\Phi}) = 18$. Moreover, when $\gamma = 0.5$, the convergence is reported to be faster as the method can converge with about $50$ iterations.

%% file: Actions/R2C5.tex
As the values of $\alpha,\epsilon,\beta$ are difficult to obtain and reveal convergence towards a stable point, which is a priori unknown, we have set the maximum number of iterations to $200$.

%% file: Actions/mentioned-ROC-monte-carlo.tex
In this part, we generate \ac{ROC} curves through Monte Carlo simulations. In particular, every point of probability of false alarm and detection probability pair  ($P_{\tt{fa}},P_{\tt{d}}$) is obtained using $10^6$ Monte Carlo trials.

%% file: Actions/Addedbenchmarks1.tex
Fig. \ref{fig:NMSE} shows the resulting \ac{NMSE} performance for the single-user case, when compared to multiple benchmarks, i.e. $(i)$ \ac{DFT} sequence used for pilot design \cite{7938362}, i.e. columns drawn out from the \ac{DFT} matrix and $(ii)$ an eigen-based design whereby the strongest $L$ eigenvectors are computed from the overall covariance matrix of the channel to construct the orthogonal pilots \cite{9499052}, and $(iii)$ a non-optimized but orthogonal pilot matrix.
We have fixed $N_t =40$ and $N_r = 70$ antennas.
The azimuth spread is set to $\bar{\sigma}_k = 4^\circ$ per user.
We have fixed the target at $\theta_0 = -20^\circ$, and set $L = 10$.
From Fig. \ref{fig:NMSE}, it is observed that the \ac{NMSE} performance of the non-optimized pilot matrix closely matches that of the pilot matrix designed using the proposed method with $\rho = 0$ (sensing-optimal configuration).
We also see in that case that the eigen-based pilot design outperforms the non-optimized pilot design by $2\dB$ and the proposed pilot for $\rho=0$ by about $1.7\dB$ gain, in terms of \ac{SNR}. Increasing $\rho$ to $0.2$ achieves a performance level comparable to the communication-optimal pilot matrix ($\rho = 1$), which provides an approximate gain of $6\dB$ over the non-optimized pilot matrix and a $4\dB$ gain over the eigen-based pilot design.
We also see that the \ac{DFT} sequence has the worst-performance as also reported in \cite{7938362}, but to further improve \ac{DFT} sequence performance in terms of channel estimation, it should be combined with range matching pursuit \cite{7938362}. 
%Analyzing the single-user case, we see that the \ac{NMSE} performance of the non-optimized pilot matrix is very close to the pilot matrix generated by the proposed method for $\rho = 0$ (sensing-optimal). Increasing $\rho$ to $0.2$ generates performance close to the communication-optimal pilot matrix ($\rho = 1$) which gives a gain of about $6\dB$ as compared to the non-optimized pilot matrix.

%For the multi-user case at $K = 4$, we can observe a more interesting phenomenon. Even though the sensing-optimal performance worse than the non-optimized pilot matrix, roughly by $2\dB$, a wider range of trade-offs can be achieved. For example, for $\rho = 0.2$, we can gain an $\NMSE$ of about $3\dB$ and again a gain of about $6\dB$ when tuning for a communication-optimal orthogonal pilot matrix, i.e. $\rho = 1$.

%% file: Actions/updated_convergence_behaviour.tex
In Fig. \ref{fig:Iter_1}, the convergence behavior of the projected gradient descent method for different values of $\rho$, in terms of $\mathcal{M}^{\tt{sense}}(\pmb{\Phi})$ and $\mathcal{M}^{\tt{comm}}(\pmb{\Phi})$ is presented.
We choose the same initial pilot for different $\rho $ to study the path taken by the proposed method.
A single-user case is analyzed, where the simulation parameters are the same as the single-user case in Fig. \ref{fig:NMSE}. The communication noise level is set to $\sigma_1 = 0.2$.
We set the azimuth spread to $\bar{\sigma}_1 = 6^\circ$.
The step-size is $\gamma = 0.1$.
The figure was generated for each $\rho$ where we store each generated pilot per iteration, and subsequently compute $\mathcal{M}^{\tt{comm}}(\pmb{\Phi})$ using \eqref{eq:M-comm-k} and $\mathcal{M}^{\tt{sense}}(\pmb{\Phi})$ via \eqref{eq:M-sense}. 
Therefore, each path corresponds to a different values of $\rho$. We have $21$ paths simulated by increasing $\rho$ a value of $0.05$ each time. 
It is evident that, for any value of $\rho$, the proposed method is ensured to converge to a fixed point. In addition, it is noteworthy that all paths eventually converge to a stable frontier. For instance, when $\rho = 0$, the method converges to an orthogonal pilot matrix whose sensing and communication \ac{MI} values are 
$(21.75,5.14)$. When $\rho = 0.5$, we converge to $(21.72,6.08)$ and $(19.72,6.38)$ when $\rho = 1$. This shows that the algorithm is able to tradeoff between sensing and communications.
% rho = 0 --> $(21.75,5.14)$ 
% rho = 0.5 --> $(21.72,6.08)$ 
% rho = 1 --> $(19.72,6.38)$ 

%% file: sections/conclusions.tex
This paper formulated a thorough \ac{ISAC} framework for designing orthogonal pilots tailored for communication and sensing purposes. 
With the help of mutual information quantities, we designed orthogonal pilots that are optimized for both communication and target detection. 
For this, a multi-objective optimization problem has been formulated to model the pilot design problem at hand. Nevertheless, a scalarization procedure was adopted, followed by an algorithmic description based on projected gradient descent was derived to generate our \ac{ISAC} orthogonal pilots.
Our analysis also reveals that the proposed design is guaranteed to converge to a stable orthogonal pilot. 
In addition, simulation results unveil the full dual potential of the design pilots, as well as the \ac{ISAC} tradeoffs the framework has to offer.\\
Future research will be oriented towards pilot generation with additional practical properties, such as good synchronization properties and low \ac{PAPR}. A possible direction may also be to utilize artificial intelligence and deep learning techniques for pilot generation, offering various features, which include but are not limited to multi-target detection and channel tracking.

%% file: sections/proof-of-Mcomm.tex
\noindent This part of the appendix re-writes the mutual information between the received signal at the $k^{th}$ communication user and the channel between the \textcolor{black}{\ac{ISAC}} \ac{BS} and the $k^{th}$ communication user. To this end, we have the following set of equalities and approximations 
% \left
% \right
\textcolor{black}{
\begin{equation}
\label{eq:MI-for-comm-1}
	\begin{split}
	&I\left(\pmb{y}_k; \pmb{h}_k\right) \\ 
	&=
	h\left(\pmb{y}_k\right)
	-
	h \left(\pmb{y}_k \vert \pmb{h}_k \right) \\
	&\myeqa 
	-\textcolor{black}{\mathbb{E}\left[\log \phi_y\left(\pmb{y}_k\right) \right]}
	-\textcolor{black}{\mathbb{E}\left[\log \phi_y\left(\pmb{y}_k \vert \pmb{h}_k\right) \right]} \\
	&\myapproxeqb
	-\textcolor{black}{ \mathbb{E} \left[ \log \phi_y\left(\bar{\pmb{y}}_k \right) + \pmb{d}^H\left(\bar{\pmb{y}}_k\right)\left[\pmb{y}_k - \bar{\pmb{y}}_k \right] \right]}
	-\textcolor{black}{\mathbb{E}\left[\log \phi_y\left(\pmb{y}_k \vert \pmb{h}_k\right) \right]} \\
	& \myeqc
	-  \log \phi_y \left(\bar{\pmb{y}}_k \right) - \pmb{d}^H \left(\bar{\pmb{y}}_k \right)  [\textcolor{black}{\mathbb{E}\left[\pmb{y}_k\right]} - \bar{\pmb{y}}_k]  
	-\textcolor{black}{\mathbb{E}\left[\log \phi_y\left(\pmb{y}_k \vert \pmb{h}_k\right) \right]} \\ 
	& \myeqd 
	-  \log \left( \sum\nolimits_{n=1}^{N_k} \alpha_{k,n} \phi_n\left(\bar{\pmb{y}}_k \right)
	  \right)   
	-\textcolor{black}{\mathbb{E}\left[\log \phi_y\left(\pmb{y}_k \vert \pmb{h}_k\right) \right]} \\ 
	& \myeqe - \log \left(\sum\nolimits_{n=1}^{N_k}  \frac{\alpha_{k,n} e^{-\bar{\pmb{\mu}}_{k,n}^H\pmb{\Phi}^H \pmb{\Sigma}_{k,n}^{-1}\left(\pmb{\Phi}\right)\pmb{\Phi}\bar{\pmb{\mu}}_{k,n}  }}{\det \left(\pmb{\Sigma}_{k,n}\left(\pmb{\Phi}\right)\right) } \right) + \tt{cnst}.
	\end{split}
\end{equation}
}
In \eqref{eq:MI-for-comm-1}, step {\sf{(a)}} follows from the definition of differential entropy and conditional differential entropy, respectively. Note that $\phi_y$ is the \ac{PDF} describing random variable $\pmb{y}_k$, i.e. $\pmb{y}_k \sim \phi_y$.
Step {\sf{(b)}} linearizes the $\log$ function through a $1^{st}$ order Taylor expansion around the mean of ${\pmb{y}}_k$, i.e. 
\begin{equation}
\label{eq:y_bar}
	\bar{\pmb{y}}_k = \textcolor{black}{\mathbb{E}[{\pmb{y}}_k]} = 
	\sum\nolimits_{n = 1}^{N_k}
	\alpha_{k,n}
	\pmb{\Phi}	
	\pmb{\mu}_{k,n},
\end{equation}
where $\pmb{d}(\pmb{y})$ is the gradient of $\log \phi_y(\pmb{y})$ with respect to $\pmb{y}$. 
Step {\sf{(c)}} applies the expectation onto random variable $\pmb{y}_k$. 
Step {\sf{(d)}} uses $\bar{\pmb{y}}_k \triangleq \textcolor{black}{\mathbb{E}[{\pmb{y}}_k]}$ and uses the fact that $\pmb{y}_k$ is also a \ac{GMM}, whose \ac{PDF} is given as
\begin{equation}
\label{eq:phi_y_at_yk}
	\phi_y(\pmb{y}_k)
	=
	\sum\nolimits_{n = 1}^{N_k}
	\alpha_{k,n}
	\phi_n(\pmb{y}_k),
\end{equation}
where it is easy to show that each Gaussian component $\phi_n$ is 
\textcolor{black}{$\phi_n(\pmb{y}_k)=f_{\mathcal{CN}}(\pmb{y}_k \vert\pmb{\Phi}\pmb{\mu}_{k,n}, \pmb{\Sigma}_{k,n}(\pmb{\Phi}))$}, and the expression of $\pmb{\Sigma}_{k,n}(\pmb{\Phi})$ is given in \eqref{eq:Sigma_Phi}.
Furthermore, we evaluate $\phi_y(\pmb{y})$ in \eqref{eq:phi_y_at_yk} at $\bar{\pmb{y}}_k$ given in \eqref{eq:y_bar} to obtain $ \phi_n(\bar{\pmb{y}}_k )$ in step {\sf{(d)}} in \eqref{eq:MI-for-comm-1}. Indeed, after some straightforward linear algebra, we can show 
 \begin{equation}
 \label{eq:PDF-of-ybar}
 	\phi_n(\bar{\pmb{y}}_k) 
 	=
  	\frac{1}{\pi^L \det ( \pmb{\Sigma}_{k,n}(\pmb{\Phi})) } \times
	\exp(-\bar{\pmb{\mu}}_{k,n}^H\pmb{\Phi}^H \pmb{\Sigma}_{k,n}^{-1}(\pmb{\Phi})\pmb{\Phi}\bar{\pmb{\mu}}_{k,n}  ),
 \end{equation}
where the expression of $\bar{\pmb{\mu}}_{k,n}$ is given in \eqref{eq:mu_bar}. Finally, injecting \eqref{eq:PDF-of-ybar} in the first term of step {\sf{(d)}} in \eqref{eq:MI-for-comm-1}, we arrive at step {\sf{(e)}}, where ${\tt{cnst}} = L \log \pi - L \log(\pi \sigma_k^2 e )$. The first term in ${\tt{cnst}}$ is due to the $\pi^L$ appearing in the denominator of \eqref{eq:PDF-of-ybar}, whereas the second term is due to the fact that the differential entropy conditioned over the channel can be written as $\textcolor{black}{\mathbb{E}\big[\log \phi_y(\pmb{y}_k \vert \pmb{h}_k) \big]} = \textcolor{black}{\mathbb{E}\big[\log \phi_y(\pmb{y}_k - \pmb{\Phi}\pmb{h}_k \vert \pmb{h}_k) \big]} = \textcolor{black}{\mathbb{E}[\phi_n(\pmb{n}_k \vert \pmb{h}_k)]} = - L \log(\pi \sigma_k^2 e )$, since $\pmb{n}_k$ is a complex Gaussian distribution.

%\begin{equation}
%	\mathcal{CN}(\pmb{\mu},\pmb{\Sigma})
%	=
%	\frac{1}{\pi^n \det(\pmb{\Sigma})}
%	\exp[-(\pmb{z} -\pmb{\mu})^H  \pmb{\Sigma}^{-1} (\pmb{z} -\pmb{\mu})]
%\end{equation}
%\begin{equation}
%\begin{split}
%	\phi_n(\bar{\pmb{y}}_k) 
%	&=
%	\frac{1}{\pi^L \det ( \pmb{\Sigma}_{k,n}(\pmb{\Phi})) } \times
%	\exp(-(	
%	\bar{\pmb{y}}_k
%	- \pmb{\Phi}\pmb{\mu}_{k,n})^H\pmb{\Sigma}_{k,n}^{-1}(\pmb{\Phi})(	\bar{\pmb{y}}_k  - \pmb{\Phi}\pmb{\mu}_{k,n}) )) \\
%	&=
%	\frac{1}{\pi^L \det ( \pmb{\Sigma}_{k,n}(\pmb{\Phi})) } \times
%	\exp(-(	
%	\sum\limits_{n' = 1}^{N_k}
%	\alpha_{k,n'}
%	\pmb{\Phi}	
%	\pmb{\mu}_{k,n'} 
%	- \pmb{\Phi}\pmb{\mu}_{k,n})^H\pmb{\Sigma}_{k,n}^{-1}(\pmb{\Phi})(	\sum\limits_{n' = 1}^{N_k}
%	\alpha_{k,n'}
%	\pmb{\Phi}	
%	\pmb{\mu}_{k,n'}  - \pmb{\Phi}\pmb{\mu}_{k,n}) )) \\
%	&=
%	\frac{1}{\pi^L \det ( \pmb{\Sigma}_{k,n}(\pmb{\Phi})) } \times
%	\exp(-(	
%	\sum\limits_{n' = 1}^{N_k}
%	\alpha_{k,n'}
%	\pmb{\mu}_{k,n'} 
%	- \pmb{\mu}_{k,n})^H\pmb{\Phi}^H \pmb{\Sigma}_{k,n}^{-1}(\pmb{\Phi})\pmb{\Phi}(	\sum\limits_{n' = 1}^{N_k}
%	\alpha_{k,n'}	
%	\pmb{\mu}_{k,n'}  - \pmb{\mu}_{k,n}) )) \\
% &= 
% 	\frac{1}{\pi^L \det ( \pmb{\Sigma}_{k,n}(\pmb{\Phi})) } \times
%	\exp(-\bar{\pmb{\mu}}_{k,n}^H\pmb{\Phi}^H \pmb{\Sigma}_{k,n}^{-1}(\pmb{\Phi})\pmb{\Phi}\bar{\pmb{\mu}}_{k,n}  ) 
%\end{split}
%\end{equation}

%% file: sections/proof-of-Msense.tex
\noindent In this appendix, we express the mutual information for sensing in terms of power quantities of the target of interest and clutter, as well as the cross-correlations between target of interest and the clutter. Therefore, we have the following set of equalities
\textcolor{black}{
\begin{equation}
\label{eq:M_sense_derivation}
	\begin{split}
		&\mathcal{M}^{\tt{sense}}\left(\pmb{\Phi} \right)  \\ 
		&\triangleq I\left(\pmb{Y}_r;\pmb{\Theta},\pmb{\nu} \vert \pmb{\Phi}\right) \myeqa h\left(\pmb{Y}_r \vert \pmb{\Phi} \right) - h\left(\pmb{Y}_r \vert \pmb{\Phi}, \pmb{\Theta},\pmb{\nu} \right) \\
		&\myeqb \log\left[\left(\pi e\right)^{N_r}\det\left\lbrace\left(\pmb{R}_{\pmb{cc}} + \sigma_r^2 \pmb{I}\right)^{-\frac{1}{2}}\pmb{R}_{\pmb{dd}}\left(\pmb{R}_{\pmb{cc}} + \sigma_r^2 \pmb{I}\right)^{-\frac{1}{2}} + \pmb{I}\right\rbrace\right] \\
		& - \log\left(\left(\pi e \right)^{N_r}\det \left\lbrace \pmb{I} \right\rbrace\right) \\
		&\myeqc \log \left[ \det\left\lbrace\left(\pmb{R}_{\pmb{cc}} + \sigma_r^2 \pmb{I}\right)^{-1}\pmb{R}_{\pmb{dd}} + \pmb{I} \right\rbrace \right] \\
		&\myeqd \log \left(1 + \nu_0\pmb{\mu}_0^H \left( \sum\nolimits_{i=1}^Q  \nu_i \pmb{\mu}_i\pmb{\mu}_i^H  + \sigma_r^2 \pmb{I} \right)^{-1} \pmb{\mu}_0 \right) 
		\\
		&\myapproxeqe \log \left(1+ \frac{\nu_0}{\sigma_r^2 } \Vert \pmb{\mu}_0 \Vert^2 -\sum\nolimits_{i=1}^Q  \frac{\frac{\nu_0}{\sigma_r^2 }\nu_i  \vert\pmb{\mu}_0^H\pmb{\mu}_i \vert^2}{\sigma_r^2 + \nu_i \Vert \pmb{\mu}_i \Vert^2}\right). \\
	\end{split}
\end{equation}
}
In step {\sf{(a)}}, we applied the definition of mutual information.
In step {\sf{(b)}}, we followed the definition of (conditional) differential entropy, and used the equivalence between \eqref{eq:Hypothesis-test-1} and \eqref{eq:Hypothesis-test-2}.
In step {\sf{(c)}}, we used $\det(\pmb{I} + \pmb{AB}) = \det(\pmb{I} + \pmb{BA})$.  
In step {\sf{(d)}}, we used $\det(\pmb{I} + \pmb{A}\pmb{x}\pmb{x}^H) = 1 + \pmb{x}^H\pmb{A}\pmb{x}$, and we expressed the covariance matrices in terms of $\pmb{\mu}_i$, for all $i= 0 \ldots Q$, i.e. $\pmb{R}_{\pmb{dd}} = \nu_0 \pmb{\mu}_0\pmb{\mu}_0^H$ and $\pmb{R}_{\pmb{cc}} = \sum_{i=1}^Q \nu_i \pmb{\mu}_i\pmb{\mu}_i^H$.   
In step {\sf{(e)}}, we have applied the Woodbury matrix identity, i.e. $(\pmb{I} + \pmb{UU}^H)^{-1} = \pmb{I} - \pmb{U} (\pmb{I} + \pmb{U}^H \pmb{U})^{-1} \pmb{U}^H $, where $\pmb{U} = \begin{bmatrix}
	\frac{\sqrt{\nu_1}}{\sigma_r} \pmb{\mu}_1 & \ldots & \frac{\sqrt{\nu_Q}}{\sigma_r} \pmb{\mu}_Q
\end{bmatrix}$ and approximated $\pmb{U}^H \pmb{U}$ to be diagonal, which is valid when $N_r$ grows large. 

%% file: sections/proof-of-dMcomm.tex
In this part of the appendix, we derive the gradient expression of $\mathcal{M}_k^{\tt{comm}}(\pmb{\Phi})$ with respect to $\pmb{\Phi}$. Using matrix-based gradient identities, we have the following set of equations
\textcolor{black}{
\begin{equation}
\begin{split}
	& \nabla_{\pmb{\Phi}}\mathcal{M}_k^{\tt{comm}}\left(\pmb{\Phi}\right) \\
	&=
	- \nabla_{\pmb{\Phi}} \log  d_{k,n}  =
	-  \omega_k^{-1}\left(\pmb{\Phi}\right)
	\sum\nolimits_{n=1}^{N_k}
	\nabla_{\pmb{\Phi}}
	d_{k,n}	 \\
	&= 
	\begin{aligned}[t]
	&-  \omega_k^{-1}\left(\pmb{\Phi}\right)
	\sum\nolimits_{n=1}^{N_k}\alpha_{k,n} 
	 \nabla_{\pmb{\Phi}}
	 \left[
	\frac{1}{\det \left(\pmb{\Sigma}_{k,n}\left(\pmb{\Phi}\right)\right) }	  \right] e^{-\beta_{k,n}(\pmb{\Phi})} \\ &
	-  \omega_k^{-1}\left(\pmb{\Phi}\right)
	\sum\nolimits_{n=1}^{N_k}\alpha_{k,n}
	\frac{1}{\det \left(\pmb{\Sigma}_{k,n}\left(\pmb{\Phi}\right)\right) }	 \nabla_{\pmb{\Phi}} e^{-\beta_{k,n}\left(\pmb{\Phi}\right)}
	 \end{aligned}\\
	&= 
	\begin{aligned}[t]
	 & \omega_k^{-1}\left(\pmb{\Phi}\right)
	\sum\nolimits_{n=1}^{N_k}\alpha_{k,n} 
	\frac{\nabla_{\pmb{\Phi}} \det\left(\pmb{\Sigma}_{k,n}\left(\pmb{\Phi}\right)\right)  }{\det^2 (\pmb{\Sigma}_{k,n}\left(\pmb{\Phi})\right) }	  e^{-\beta_{k,n}\left(\pmb{\Phi}\right)} \\ &
	+\omega_k^{-1}\left(\pmb{\Phi}\right)
\sum\nolimits_{n=1}^{N_k}\alpha_{k,n} 
	\frac{\nabla_{\pmb{\Phi}} \beta_{k,n}\left(\pmb{\Phi}\right)}{\det \left(\pmb{\Sigma}_{k,n}\left(\pmb{\Phi}\right)\right) }	  e^{-\beta_{k,n}\left(\pmb{\Phi}\right)}
	 \end{aligned}\\
	 &= \omega_k^{-1}\left(\pmb{\Phi}\right)
	 \sum\nolimits_{n=1}^{N_k}
	 d_{k,n}
	  \left\lbrace
	 \frac{\nabla_{\pmb{\Phi}}\det \left(\pmb{\Sigma}_{k,n}\left(\pmb{\Phi}\right)\right)}{\det \left(\pmb{\Sigma}_{k,n}\left(\pmb{\Phi}\right)\right)}
	 +
	 \nabla_{\pmb{\Phi}} \beta_{k,n}\left(\pmb{\Phi}\right)
	  \right\rbrace \\
	 &= \omega_k^{-1}\left(\pmb{\Phi}\right)
	 \sum\nolimits_{n=1}^{N_k}
	 d_{k,n}
	  \left\lbrace
	 \pmb{\Sigma}^{-1}_{k,n}\left(\pmb{\Phi}\right)
	 \nabla_{\pmb{\Phi}}
	 \pmb{\Sigma}_{k,n}\left(\pmb{\Phi}\right)
	 +
	 \nabla_{\pmb{\Phi}} \beta_{k,n}\left(\pmb{\Phi}\right)
	  \right\rbrace \\
	 &= 		\begin{aligned}[t]
& 2\omega_k^{-1}\left(\pmb{\Phi}\right)
	 \sum\nolimits_{n=1}^{N_k}
	 d_{k,n}
	 \pmb{\Sigma}^{-1}_{k,n}\left(\pmb{\Phi}\right)
	 \pmb{\Phi}
	 \pmb{R}_{k,n} 
	  \\ & 
-	2\omega_k^{-1}\left(\pmb{\Phi}\right)
	 \sum\nolimits_{n=1}^{N_k}
	 d_{k,n}\pmb{R}_{k,n}\pmb{\Phi}^H \pmb{\Sigma}^{-1}_{k,n}\left(\pmb{\Phi}\right) \pmb{\Phi} \bar{\pmb{\mu}}_{k,n}\bar{\pmb{\mu}}_{k,n}^H \pmb{\Phi}^H  \pmb{\Sigma}^{-1}_{k,n}\left(\pmb{\Phi}\right) \\ &
+ 	2\omega_k^{-1}\left(\pmb{\Phi}\right)
	 \sum\nolimits_{n=1}^{N_k}
	 d_{k,n} \bar{\pmb{\mu}}_{k,n}\bar{\pmb{\mu}}_{k,n}^H\pmb{\Phi}^H \pmb{\Sigma}^{-1}_{k,n}\left(\pmb{\Phi}\right) ,
	 \end{aligned}
\end{split}
\end{equation}
}
where $\omega_k(\pmb{\Phi}) = \sum_{n=1}^{N_k}  d_{k,n}$ and $d_{k,n} = \frac{\alpha_{k,n}e^{-\beta_{k,n}(\pmb{\Phi})} }{\det (\pmb{\Sigma}_{k,n}(\pmb{\Phi}))}$ have been introduced for sake of compact notation.

%% file: sections/proof-of-dMsense.tex
In this part of the appendix, we derive the gradient of $\mathcal{M}^{\tt{sense}}(\pmb{\Phi}) $ with respect to the $l^{th}$ pilot vector $\pmb{\Phi}_l$. Due to the $\log$ structure, it is easily verified that we can write
\begin{equation}
\label{eq:first-gradient-dMsense}
	\nabla_{\pmb{\Phi}_l}\mathcal{M}^{\tt{sense}}(\pmb{\Phi}) =  
	\frac{\frac{\nu_0}{\sigma_r^2 } \nabla_{\pmb{\Phi}_l}\Vert \pmb{\mu}_0 \Vert^2 -
	\frac{\nu_0}{\sigma_r^2 }\sum\nolimits_{i=1}^Q  \nabla_{\pmb{\Phi}_l}\frac{  \nu_i\vert\pmb{\mu}_0^H\pmb{\mu}_i \vert^2}{\sigma_r^2 + \nu_i \Vert \pmb{\mu}_i \Vert^2}}{1+ \frac{\nu_0}{\sigma_r^2 } \Vert \pmb{\mu}_0 \Vert^2 -
	\frac{\nu_0}{\sigma_r^2 }\sum\nolimits_{i=1}^Q  \frac{  \nu_i\vert\pmb{\mu}_0^H\pmb{\mu}_i \vert^2}{\sigma_r^2 + \nu_i \Vert \pmb{\mu}_i \Vert^2}}.
\end{equation}
Using the following expressions
\begin{align}
	 \nabla_{\pmb{\Phi}_l}\Vert \pmb{\mu}_i \Vert^2
	 &=
	 2 \pmb{A}^H(\theta_i)\pmb{A}(\theta_i)\pmb{\Phi}_l, 	 \\
	 \nabla_{\pmb{\Phi}_l}\pmb{\mu}_i^H \pmb{\mu}_j
	 &=
	 \pmb{A}^H(\theta_i)\pmb{A}(\theta_j)\pmb{\Phi}_l 
	 +
	 \pmb{A}^H(\theta_j)\pmb{A}(\theta_i)\pmb{\Phi}_l, \\
	 \nabla_{\pmb{\Phi}_l} \vert \pmb{\mu}_i^H \pmb{\mu}_j\vert^2 &= 
	 2\Re[\pmb{\mu}_i^H \pmb{\mu}_j] \big( \pmb{A}^H(\theta_i)\pmb{A}(\theta_j) + \pmb{A}^H(\theta_j)\pmb{A}(\theta_i) \big)\pmb{\Phi}_l,
\end{align}
we can express equation \eqref{eq:first-gradient-dMsense} as follows
%\begin{equation}
%	\nabla_{\pmb{\Phi}_l}\mathcal{M}^{\tt{sense}}(\pmb{\Phi}) =  
%	\frac{\frac{2\nu_0}{\sigma_r^2 }\pmb{A}^H(\theta_i)\pmb{A}(\theta_i)\pmb{\Phi}_l -
%	\frac{\nu_0}{\sigma_r^2 }\sum\limits_{i=1}^Q  \nabla_{\pmb{\Phi}_l}\frac{  \nu_i\vert\pmb{\mu}_0^H\pmb{\mu}_i \vert^2}{\sigma_r^2 + \nu_i \Vert \pmb{\mu}_i \Vert^2}}{1+ \frac{\nu_0}{\sigma_r^2 } \Vert \pmb{\mu}_0 \Vert^2 -
%	\frac{\nu_0}{\sigma_r^2 }\sum\limits_{i=1}^Q  \frac{  \nu_i\vert\pmb{\mu}_0^H\pmb{\mu}_i \vert^2}{\sigma_r^2 + \nu_i \Vert \pmb{\mu}_i \Vert^2}}.
%\end{equation}
\begin{equation}
	\nabla_{\pmb{\Phi}_l}\mathcal{M}^{\tt{sense}}(\pmb{\Phi}) =  
	\frac{g_0(\pmb{\Phi}) - \sum_{i=1}^Q g_i(\pmb{\Phi})}{1+ \frac{\nu_0}{\sigma_r^2 } \Vert \pmb{\mu}_0 \Vert^2 -
	\frac{\nu_0}{\sigma_r^2 }\sum\nolimits_{i=1}^Q  \frac{  \nu_i\vert\pmb{\mu}_0^H\pmb{\mu}_i \vert^2}{\sigma_r^2 + \nu_i \Vert \pmb{\mu}_i \Vert^2}}.
\end{equation}
where $g_i = g_i^{(1)} + g_i^{(2)}$ and
\begin{align}
	g_0(\pmb{\Phi}) &= \frac{2\nu_0}{\sigma_r^2 }\pmb{A}^H(\theta_i)\pmb{A}(\theta_i)\pmb{\Phi}_l, \\
	g_i^{(1)}(\pmb{\Phi}) &= \frac{2\nu_i \Re[\pmb{\mu}_0^H \pmb{\mu}_i]}{(\sigma_r^2 + \nu_i \Vert \pmb{\mu}_i \Vert^2)}\big( \pmb{A}^H(\theta_0)\pmb{A}(\theta_i) + \pmb{A}^H(\theta_i)\pmb{A}(\theta_0) \big)\pmb{\Phi}_l, \\
	g_i^{(2)}(\pmb{\Phi}) &= - \frac{2\nu_i^2 \vert\pmb{\mu}_0^H\pmb{\mu}_i \vert^2 }{(\sigma_r^2 + \nu_i \Vert \pmb{\mu}_i \Vert^2)^2}\pmb{A}^H(\theta_i)\pmb{A}(\theta_i)\pmb{\Phi}_l.
\end{align}

%% file: sections/convergence-analysis-proof.tex
Before demonstrating the proof, we introduce the following  real-valued vectorial notation for compactness. We first denote $f(\pmb{\phi}) = -\mathcal{M}^{\tt{ISAC}}(\pmb{\Phi})$, \textcolor{black}{\input{Actions/C3o3_phidefinition}}As a result, we can reformulate projected gradient descent as follows
\begin{equation}
	\nabla 
	f(\pmb{\phi})
	=
	-
	\begin{bmatrix}
		\Real(\ve(\nabla \mathcal{M}^{\tt{ISAC}}(\pmb{\Phi}))) \\
		\Imag(\ve(\nabla \mathcal{M}^{\tt{ISAC}}(\pmb{\Phi}))) \\
	\end{bmatrix}.
\end{equation}
The update equations are now simply represented as 
\begin{equation}
\label{eq:PGD-vector-real}
	\pmb{z}_{t+1} 
	=
	\pmb{\phi}_t
	-
	\gamma
	\nabla
	f(\pmb{\phi}_t),
\end{equation}
where $\pmb{\phi}_{t+1}$ is further vectorized as follows
\begin{equation}
	\label{eq:PGD-vector-project}
	\pmb{\phi}_{t+1}
	\triangleq
	\bar{\pmb{\pi}}(\pmb{Z}_{t+1})
	=
	-
	\begin{bmatrix}
		\Real(\ve(\pmb{\pi}(\pmb{Z}_{t+1} ))) \\
		\Imag(\ve(\pmb{\pi}(\pmb{Z}_{t+1} ))) \\
	\end{bmatrix}.
\end{equation}
Using the \ac{RSS} condition in \textbf{Definition 1}, and exploiting \eqref{eq:RSS} for $\pmb{x}_1 = \pmb{\phi}_t$ and $\pmb{x}_2 = \pmb{\phi}_{t+1}$, we have the following series of (in)equalities that hold true for all $ \pmb{\phi}_t, \pmb{\phi}_{t+1}$, namely
\begin{equation}
\label{eq:RSS-bound}
	\begin{split}
\Delta_{t+1} 
&\myleqa \nabla f^T(\pmb{\phi}_{t})(\pmb{\phi}_{t+1}-\pmb{\phi}_{t}) + \frac{\beta}{2} \Vert \pmb{\phi}_{t} - \pmb{\phi}_{t+1} \Vert^2 + \frac{\alpha}{2} \epsilon^2 \\
&\myeqb \frac{1}{\gamma}
(
	\pmb{\phi}_t
		-
	\pmb{z}_{t+1} 
)^T(\pmb{\phi}_{t+1}-\pmb{\phi}_{t}) + \frac{\beta}{2} \Vert \pmb{\phi}_{t} - \pmb{\phi}_{t+1} \Vert^2 + \frac{\alpha}{2} \epsilon^2 \\
&\myeqc \frac{\beta}{2}
(
\Vert  \pmb{\phi}_{t+1} - \pmb{z}_{t+1} \Vert^2
-
\Vert  \pmb{\phi}_{t}  - \pmb{z}_{t+1} \Vert^2
) 
+ \frac{\alpha}{2} \epsilon^2 \\ 
&\myleqd  \frac{\beta}{2}
(
\Vert  \pmb{\phi}^{{\tt{opt}}} - \pmb{z}_{t+1} \Vert^2
-
\Vert  \pmb{\phi}_{t}  - \pmb{z}_{t+1} \Vert^2
) 
+ \frac{\alpha}{2} \epsilon^2 \\ 
& \myeqe
\frac{\beta}{2}
(\Vert  \pmb{\phi}^{{\tt{opt}}} - \pmb{\phi}_{t} \Vert^2
+
2(\pmb{\phi}^{{\tt{opt}}} - \pmb{\phi}_{t} )^T(\pmb{\phi}_{t}  - \pmb{z}_{t+1})
)
+ \frac{\alpha}{2} \epsilon^2 \\
& \myeqf
\frac{\beta}{2}
\Vert  \pmb{\phi}^{{\tt{opt}}} - \pmb{\phi}_{t} \Vert^2
+
(\pmb{\phi}^{{\tt{opt}}} - \pmb{\phi}_{t} )^T\nabla f(\pmb{\phi}_{t})
+ \frac{\alpha}{2} \epsilon^2,
	\end{split}
\end{equation}
where $ \Delta_{t+1} = f(\pmb{\phi}_{t+1}) - f(\pmb{\phi}_{t})$ is the drift in cost at iteration $t+1$ of the projected gradient descent algorithm.
In step {\sf{(a)}}, we utilized the \textbf{Definition 1} at $ \pmb{\phi}_t, \pmb{\phi}_{t+1}$.
In step {\sf{(b)}}, we used \eqref{eq:PGD-vector-real}.
In step {\sf{(c)}}, we used a descent step size of $\gamma = \frac{1}{\beta}$ and the identity 
\begin{equation}
\label{eq:norm-identity}
	(\pmb{a}-\pmb{b})^T(\pmb{b}-\pmb{c}) = \frac{1}{2}(\Vert \pmb{a} - \pmb{c} \Vert^2 - \Vert \pmb{a} - \pmb{b} \Vert^2 - \Vert \pmb{b} - \pmb{c} \Vert^2),
\end{equation}
for any real-valued $p-$dimensional vectors, $\pmb{a},\pmb{b},\pmb{c}$.
In step {\sf{(d)}}, we have used the property that for all \textcolor{black}{$\pmb{x} = \bar{\pmb{\pi}}(\pmb{X})$}, such that $\pmb{X} \in {\tt{St}}(L,N_t)$, we have that $\Vert \pmb{\phi} - \pmb{z} \Vert^2 \leq \Vert \pmb{x} - \pmb{z} \Vert^2$, for $\pmb{\phi} = \bar{\pmb{\pi}}(\pmb{z})$ and $\bar{\pmb{\pi}}()$ is defined in \eqref{eq:PGD-vector-project}. This property is trivial, as $\pmb{\phi}$ minimizes the distance in criterion \eqref{eq:GD-project} on the Stiefel manifold. 
In step {\sf{(e)}}, we again apply identity \eqref{eq:norm-identity}.
In step {\sf{(f)}}, we used \eqref{eq:PGD-vector-real}. 

Now we leverage the \ac{RSC} condition given in \textbf{Definition 2}. This is done by applying \eqref{eq:RSC} twice. The first time through $\pmb{x}_2 = \pmb{\phi}^{{\tt{opt}}}$ and $\pmb{x}_1 = \pmb{\phi}_t$, which gives
\begin{equation}
\label{eq:RSC-first-time}
	f(\pmb{\phi}^{{\tt{opt}}}) - f(\pmb{\phi}_t) 
	\geq \nabla f^T(\pmb{\phi}_t)(\pmb{\phi}^{{\tt{opt}}}-\pmb{\phi}_t)  + \frac{\alpha}{2} \Vert \pmb{\phi}_t - \pmb{\phi}^{{\tt{opt}}} \Vert^2 - \frac{\alpha}{2} \epsilon^2,
\end{equation}
and the second through $\pmb{x}_2 = \pmb{\phi}_t$ and $\pmb{x}_1 = \pmb{\phi}^{{\tt{opt}}}$, i.e.
 \begin{equation}
 \label{eq:RSC-second-time}
	 f(\pmb{\phi}_t) - f(\pmb{\phi}^{{\tt{opt}}})
	\geq \frac{\alpha}{2} \Vert \pmb{\phi}_t - \pmb{\phi}^{{\tt{opt}}} \Vert^2 - \frac{\alpha}{2} \epsilon^2.
\end{equation}
where in $\nabla f(\pmb{\phi}^{{\tt{opt}}}) = \pmb{0}$, since the gradient is null at the optimal pilot matrix. Now, multiplying inequality \eqref{eq:RSC-second-time} by $\frac{\beta}{\alpha}$ then adding it with inequality \eqref{eq:RSC-first-time} gives
\begin{equation}
\label{eq:RSC-bound}
\begin{split}
	(1 - \frac{\beta}{\alpha})(f(\pmb{\phi}^{{\tt{opt}}}) - f(\pmb{\phi}_t) ) 
	& \geq
\frac{\beta}{2}
\Vert  \pmb{\phi}^{{\tt{opt}}}  - \pmb{\phi}_{t} \Vert^2
\\ & +
(\pmb{\phi}^{{\tt{opt}}} - \pmb{\phi}_{t} )^T\nabla f(\pmb{\phi}_{t})
-(\alpha + \beta)\frac{\epsilon^2}{2}.	
\end{split}
\end{equation}
In \eqref{eq:RSC-bound}, we have further relaxed the lower bound by removing the non-negative term, $\frac{\alpha}{2} \Vert \pmb{\phi}_t - \pmb{\phi}^{{\tt{opt}}} \Vert^2.$ Combining \eqref{eq:RSS-bound} with \eqref{eq:RSC-bound} gives
\begin{equation}
\label{eq:comb-1}
	(1 - \frac{\beta}{\alpha})(f(\pmb{\phi}^{{\tt{opt}}}) - f(\pmb{\phi}_t) ) 
	 \geq -(\alpha + \frac{\beta}{2})\epsilon^2 + \Delta_{t+1}.
\end{equation}
Then, by adding and subtracting the term $f(\pmb{\phi}^{{\tt{opt}}})$ on the lower bound in \eqref{eq:comb-1} gives
\begin{equation}
\label{eq:comb-2}
		\frac{\beta}{\alpha} (  f(\pmb{\phi}_t) - f(\pmb{\phi}^{{\tt{opt}}})) 
		+
		(\alpha + \frac{\beta}{2})\epsilon^2
	 \geq f(\pmb{\phi}_{t+1}) - f(\pmb{\phi}^{{\tt{opt}}}).
\end{equation}
Applying \eqref{eq:comb-2} over $t$ iterations enables us to bound $f(\pmb{\phi}_{t+1}) - f(\pmb{\phi}^{{\tt{opt}}})$ as follows
\begin{equation}
\begin{split}
	f(\pmb{\phi}_{t+1}) - f(\pmb{\phi}^{{\tt{opt}}}) 
	&\leq
	(\frac{\beta}{\alpha})^{t+1}
	[f(\pmb{\phi}_{0}) - f(\pmb{\phi}^{{\tt{opt}}}) ]
	\\ &+
	(\alpha + \frac{\beta}{2})\epsilon^2
	\sum\nolimits_{k=0}^t
	(\frac{\beta}{\alpha})^k,
\end{split}
\end{equation}
where by applying geometric progression, i.e. $\sum\nolimits_{k=1}^n aq^{k-1} = \frac{a(q^n - 1)}{q-1}$ \cite{gradshteyn2014table}, we get
\begin{equation}
\begin{split}
	f(\pmb{\phi}_{t+1}) - f(\pmb{\phi}^{{\tt{opt}}}) 
	&\leq
	(\frac{\beta}{\alpha})^{t+1}
	(f(\pmb{\phi}_{0}) - f(\pmb{\phi}^{{\tt{opt}}}) )
	\\ &+
	(\alpha + \frac{\beta}{2})\epsilon^2
	\frac{1-(\frac{\beta}{\alpha})^{t+1} }{1-(\frac{\beta}{\alpha})}.
\end{split}
\end{equation}
Provided that $\frac{\beta}{\alpha} < 1$, it is easily observed that as $t \rightarrow \infty$, we have that
\begin{equation}
	f(\pmb{\phi}_{\infty}) - f(\pmb{\phi}^{{\tt{opt}}}) 
	\leq
	(\alpha + \frac{\beta}{2})\epsilon^2
	\frac{1}{1-\frac{\beta}{\alpha}}.
\end{equation}
Replacing $f(\pmb{\phi})$ with $-\mathcal{M}^{\tt{ISAC}}(\pmb{\Phi})$, we arrive at the result in \eqref{eq:convergence-bound}, which finalizes the proof of the theorem.

%% file: Actions/C3o3_phidefinition.tex
where $\pmb{\phi} = \operatorname{vec}(\pmb{\Phi})$.

%% file: Actions/lower-bound-on-Mcomm.tex
Herein, we lower bound the adopted \ac{MI} communication metric as follows,
\begin{equation}
\label{eq:MI-for-comm-2}
	\begin{split}
	\mathcal{M}_k^{\tt{comm}}(\pmb{\Phi})
	 &\mygeqa
	I(\widehat{\pmb{h}}_k;\pmb{h}_k \vert \pmb{\Phi}) \\
	& \myeqb h(\pmb{h}_k \vert \pmb{\Phi})-h(\pmb{h}_k\mid \widehat{\pmb{h}}_k, \pmb{\Phi}) \\
	& \myeqc h(\pmb{h}_k \vert \pmb{\Phi})-h(\widetilde{\boldsymbol{h}}_k \mid \widehat{\pmb{h}}_k,\pmb{\Phi}) \\
& \mygeqd h(\boldsymbol{h}_k\vert \pmb{\Phi})-h(\widetilde{\boldsymbol{h}}_k\vert \pmb{\Phi}) \\
& \mygeqe h(\pmb{h}_k\vert \pmb{\Phi})-\log \left(\det \left( \pi e\boldsymbol{R}_{\widetilde{\boldsymbol{h}}_k}\right)\right) \\
&\mygeqf h(\pmb{h}_k\vert \pmb{\Phi})-N_t \log(\pi e) -  N_t\log \frac{\operatorname{tr}(\boldsymbol{R}_{\widetilde{\boldsymbol{h}}_k})}{N_t}.
	\end{split}
\end{equation}
In the above series of steps, step (a) follows by the data processing inequality by the following Markov chain $\pmb{h}_k \rightarrow \pmb{y}_k \rightarrow \widehat{\pmb{h}}_k$ \cite{cover1999elements}.
Step (b) applies the definition of mutual information as function of entropy $h$ \cite{cover1999elements}.
Step (c) exploits the fact that adding/subtracting a conditioned quantity does not alter entropy and introduces the estimation error vector $\widetilde{\boldsymbol{h}}_k = \pmb{h}_k - \widehat{\pmb{h}}_k$.
Step (d) leverages the fact that conditioning decreases uncertainty, hence increases entropy.
Step (e) lower bounds the mutual information using the \ac{MSE} matrix, i.e. \textcolor{black}{$\boldsymbol{R}_{\widetilde{\boldsymbol{h}}_k}=$ $\mathbb{E}\left[\widetilde{\boldsymbol{h}}_k \widetilde{\boldsymbol{h}}_k^H\right]$}, by using the fact that the entropy on a Gaussian distributed random variable maximizes entropy \cite{cover1999elements}. 
Step (f) follows by applying the \textcolor{black}{arithmetic mean-geometric mean} inequality on positive semidefinite matrices, namely $\frac{\operatorname{tr}(\pmb{X})}{N} \geq(\operatorname{det}(\pmb{X}))^{\frac{1}{N}}$ where $\pmb{X} \succeq \pmb{0}$ is $N \times N$.

%% file: Actions/stein-on-m-sense.tex
For the hypothesis testing problem in \eqref{eq:Hypothesis-test-2}, we can define the corresponding probability densities per hypothesis as
\begin{equation}
	\begin{aligned}
& f\left(\boldsymbol{z} \mid \mathcal{H}_0\right) \propto \frac{1}{\left|\boldsymbol{I}\right|} e^{-\boldsymbol{z}^H \boldsymbol{I}^{-1} \boldsymbol{z}}, \\
& f\left(\boldsymbol{z} \mid \mathcal{H}_1\right) \propto \frac{1}{\left|\boldsymbol{I}+\boldsymbol{A}\right|} e^{-\boldsymbol{z}^H\left(\boldsymbol{I}+\boldsymbol{A}\right)^{-1} \boldsymbol{z}},
\end{aligned}
\end{equation}
where $\boldsymbol{A}= (\pmb{R}_{\pmb{cc}} + \sigma_r^2 \pmb{I})^{-\frac{1}{2}}\pmb{R}_{\pmb{dd}}(\pmb{R}_{\pmb{cc}} + \sigma_r^2 \pmb{I})^{-\frac{1}{2}}$.
Stein's lemma states that for a fixed probability of false alarm, we have that
\begin{equation*}
	\mathcal{D}\left(f\left(\boldsymbol{z} \mid \mathcal{H}_0\right) \| f\left(\boldsymbol{z} \mid \mathcal{H}_1\right)\right)=\lim _{L \rightarrow \infty}\left(-\frac{1}{L} \log \left(1-P_{\mathrm{D}}\right)\right).
\end{equation*}
The \ac{KL}-divergence is given as \cite{5200973}
\begin{equation}
	\mathcal{D}\left(p_0 \| p_1\right)=\textcolor{black}{\mathbb{E}_{p_0}\left[\ln \left[\frac{p_0(z)}{p_1(z)}\right]\right]}.
\end{equation}
It is worth noting that the divergence is also referred to as the power exponent of the most powerful test, i.e. the likelihood ratio test according to Neyman–Pearson lemma \cite{scharf1991statistical}.
Note that the power exponent can be expressed as 
\begin{equation}
	\begin{split}
		&\mathcal{D}\left(f\left(\boldsymbol{z} \mid \mathcal{H}_0\right) \| f\left(\boldsymbol{z} \mid \mathcal{H}_1\right)\right) \\&\myeqa \textcolor{black}{\mathbb{E}_{p_0}\left[\ln \left[\frac{|\boldsymbol{I}+\boldsymbol{A}|}{|\boldsymbol{I}|} e^{-\boldsymbol{z}^H\left(\boldsymbol{I}^{-1}-(\boldsymbol{I}+\boldsymbol{A})^{-1}\right) \boldsymbol{z}} \right]\right]} \\
		&\myeqb  \textcolor{black}{\mathbb{E}_{p_0}\left[\ln \frac{|\boldsymbol{I}+\boldsymbol{A}|}{|\boldsymbol{I}|}-\boldsymbol{z}^H\left(\boldsymbol{I}-(\boldsymbol{I}+\boldsymbol{A})^{-1}\right) \boldsymbol{z}\right]} \\
		&\myeqc  \mathbb{E}_{p_0}\left[\ln \frac{|\boldsymbol{I}+\boldsymbol{A}|}{|\boldsymbol{I}|}\right]-\mathbb{E}_{p_0}\left[\boldsymbol{z}^H\left(\boldsymbol{I}-(\boldsymbol{I}+\boldsymbol{A})^{-1}\right) \boldsymbol{z}\right] \\
		&\myeqd \ln \frac{|\boldsymbol{I}+\boldsymbol{A}|}{|\boldsymbol{I}|} - \operatorname{Tr}\left(\boldsymbol{I}-(\boldsymbol{I}+\boldsymbol{A})^{-1}\right) \\
		&\myeqe \mathcal{M}^{\tt{sense}}(\pmb{\Phi}) - \frac{\nu_0 \boldsymbol{\mu}_0^H\left(\sum_{i=1}^Q \nu_i \boldsymbol{\mu}_i \boldsymbol{\mu}_i^H+\sigma_r^2 \boldsymbol{I}\right)^{-1} \boldsymbol{\mu}_0}{1+\nu_0 \boldsymbol{\mu}_0^H\left(\sum_{i=1}^Q \nu_i \boldsymbol{\mu}_i \boldsymbol{\mu}_i^H+\sigma_r^2 \boldsymbol{I}\right)^{-1} \boldsymbol{\mu}_0}.
	\end{split}
\end{equation}
In the above series of steps, 
step (a) applies the definition of \ac{KL}-divergence.
Step (b) applies the logarithm. 
Step (c) invokes the linearity of the expectation operator.
Step (d) applies the expectation on both terms, where the first term is deterministic and the second term uses the fact that for a complex standard Gaussian vector (which is the distribution of $\boldsymbol{z}$ under $\mathcal{H}_0$), the quadratic expectation is $\mathbb{E}_{p_0}\left[\boldsymbol{z}^H \boldsymbol{M} \boldsymbol{z}\right]=\operatorname{Tr}(\boldsymbol{M})$.
Step (e) uses the definition of $\mathcal{M}^{\tt{sense}}(\pmb{\Phi})$ (equation \eqref{eq:M_sense_derivation}) on the first term, along with the Sherman-Morrison formula applied on the second term.